\documentclass[preprint]{aastex61}

\newcommand\aastex{AAS\TeX}

\accepted{on Oct 13, 2019.  The Astronomical Journal, in press}

\shorttitle{\aastex\ A Fireball and PHA (164121) 2003~YT$_1$}
\shortauthors{Kasuga et al.}

\usepackage{mediabb}
\usepackage{amsmath}
\usepackage{amssymb}
\usepackage{rotating, graphicx}
\usepackage{longtable}
\usepackage[stable]{footmisc}

\begin{document}

\title{A Fireball and Potentially Hazardous Binary Near-Earth Asteroid (164121) 2003~YT$_1$}

\correspondingauthor{Toshihiro Kasuga}
\email{toshi.kasuga@nao.ac.jp}

\author[0000-0001-5903-7391]{Toshihiro Kasuga}
\affil{National Astronomical Observatory of Japan, 2-21-1 Osawa, Mitaka, Tokyo 181-8588, Japan}
\affil{Department of Physics, Faculty of Science, Kyoto Sangyo University, Motoyama, Kamigamo, Kita-ku, Kyoto 603-8555, Japan}

\author{Mikiya Sato}
\affiliation{The Nippon Meteor Society, Japan}

\author{Masayoshi Ueda}
\affil{The Nippon Meteor Society, Japan}

\author{Yasunori Fujiwara}
\affiliation{The Nippon Meteor Society, Japan}

\author{Chie Tsuchiya}
\affil{National Astronomical Observatory of Japan, 2-21-1 Osawa, Mitaka, Tokyo 181-8588, Japan}

\author{Jun-ichi Watanabe}
\affil{National Astronomical Observatory of Japan, 2-21-1 Osawa, Mitaka, Tokyo 181-8588, Japan}



\begin{abstract}

We present a fireball detected in the night sky over Kyoto, 
Japan on UT 2017 April 28 at ${\rm 15^{h}\,58^{m}\,19^{s}}$ 
by the SonotaCo Network. 
The absolute visual magnitude is $M_{\rm v}$=$-$4.10$\pm$0.42\,mag.  
Luminous light curves obtain 
a meteoroid mass $m$=29$\pm$1\,g, 
corresponding to the size $a_{\rm s}$=2.7$\pm$0.1\,cm.  
Orbital similarity assessed by D-criterions (cf. $D_{\rm SH}$=0.0079) has 
identified a likely parent, the binary near-Earth asteroid (164121) 2003~YT$_1$.   
The suggested binary formation process is  
a YORP-driven rotational disintegration \citep[][]{Pravec07Icar}. 
The asynchronous state indicates the age of $<$\,10$^4$\,yr, 
near or shorter than the upper limit to meteoroid stream lifetime.    
We examine potential 
dust production mechanisms for the asteroid, 
including rotational instability, resurfacing, 
impact, photoionization, 
radiation pressure sweeping, 
thermal fracture and sublimation of ice.  
We find some of them capable of producing the meteoroid-scale particles.    
Rotational instability is presumed to cause mass shedding, 
in consideration of the recent precedents (e.g. asteroid (6478) Gault), 
possibly releasing mm-cm scale dust particles.
Impacts by micrometeorites with size $\simeq$\,1\,mm 
could be a trigger for ejecting the cm-sized particles.  
Radiation pressure can sweep out the mm-sized dust particles, 
while not sufficient for the cm-sized. 
For the other mechanisms, unprovable or unidentified.  
The feasibility in the parental aspect of 2003~YT$_1$ is somewhat 
reconciled with the fireball observation, yielding an insight into how we approach 
potentially hazardous objects.

\end{abstract}




\section{Introduction} 
\label{intro}

The worldwide meteor survey networks have established 
the procedure for identifying meteoroid orbits in streams 
and associated parent bodies, 
asteroids and comets, mostly known as near-Earth objects (NEOs) 
\citep{So09,RudawskaJenniskens2014,Ye2016,Jenniskens17}. 
Some NEOs, meteorite falls and fireballs have been linked with  
potentially hazardous asteroids (PHAs) \citep{Madiedo13MNRAS, Madiedo14MNRAS,Svetsov19} 
of which the Taurids are studied in many cases \citep[][]{BMM13,Olech17,SBM17,Clark2019}. 
Physical disintegration of NEOs 
result in producing orbit-hugging dust (streams) 
which may cross the Earth orbit.   
Suggested mechanisms, 
especially for those of asteroids, 
include rotational instability, thermal stress, collisions (impacts)  
and so on \citep{Jewitt12,JHA15}.  
Asteroidal stream parents should be, or used to be losing mass, 
while among the few mass-loss activities other than activity driven by sublimation of ice are identified \citep{KasugaJewitt19}.  

A relatively slow, bright fireball was detected in the sky over Kyoto, 
Japan on UT 2017 April 28 at ${\rm 15^{h}\,58^{m}\,19^{s}}$ 
through the SonotaCo Network \citep{So09}.   
The small semimajor axis ($a$=1.111\,AU) and high inclination ($i$ = 43.9$^{\circ}$) 
present its peculiar orbit.  
The dynamical properties, 
as given by orbit-linking $D$-criterions \citep[cf. ][]{SouthHawkins63}, 
find a close association with 
the near-Earth asteroid (NEA) (164121) 2003~YT$_1$ (hereafter, 2003~YT$_1$)
(See details in Section~\ref{results}).  
The short distance from the asteroid orbit to the Earth orbit (cf. 0.0026\,AU at the descending node) 
is compatible with those of meteoroid streams for showers \citep[$\lesssim$\,0.01\,AU,][]{Vaubaillon19}, 
suggesting that both the fireball and 2003~YT$_1$ practically cross the Earth orbit.  
This asteroid-meteor pair is likely to be secured, giving a rare opportunity for understanding of meteoroid production.

The NEA 2003~YT$_1$ was discovered on UT 2003 December 18 
in the course of the Catalina Sky Survey \citep{Tichy03}.  
Based upon the absolute magnitude $H$=16.2 and 
the low Minimum Orbit Intersection Distance (MOID) 
$\sim$ 0.003\,AU (NASA/JPL Small-Body Database), 
the object is a PHA \citep{Larson04,Hicks09}.  
The impact probability to the Earth is calculated $\sim$6\% per 10$^7$\,yr \citep{Galiazzo17}.  
The Arecibo radar delay-Doppler and optical photometric observations 
independently identified 2003~YT$_1$ has a binary system \citep{Nolan04IAU}.   
Suggested formation process is a 
rotational instability,  a breakup/fission driven by 
Yarkovsky-O'Keefe-Radzievskii-Paddack (YORP) torques \citep[][]{Pravec07Icar}.  
The primary has 1.1$\pm$0.2\,km in diameter ($D_{\rm p}$) 
and the secondary with a diameter of 0.21$\pm$0.06\,km ($D_{\rm s}$), 
having a distance of 2.7\,km \citep{Nolan04}.  
The primary's rotation period is 2.343$\pm$0.001\,hr, 
and the light curve amplitude of $\sim$0.16\,mag exhibits 
its nearly spheroidal shape \citep{Galad04IAU,Larson04,Warner18PDS}.  
The secondary's rotation period $\lesssim$ 6\,hr and 
its orbital period of $\sim$30\,hr (eccentric orbit) 
suggest the asynchronous state \citep{Nolan04IAU,Nolan04}.  
Geometric albedos (in visual and infrared) are measured 
by thermal infrared observations, 
$p_{\rm v}$ = 0.24$\pm$0.16 from the ground-based \citep{Delbo11},  
and $p_{\rm v}$ = 0.20$\pm$0.10 and $p_{\rm IR}$ = 0.33$\pm$0.14 
from the space \citep[{\it WISE/NEOWISE},][]{Mainzer12}.
Near-infrared spectra (0.7--2.5$\micron$) reveal the surface assemblage 
dominated by orthopyroxene with any lack of olivine content on 2003~YT$_1$, 
implying taxonomically V-type asteroid \citep[][]{AbellFahhey04,AbellGaffey05,Sanchez13}.  
The regolith breccia \citep[$<$ 25$\micron$ in size mostly,][]{Ieva16}  
could be originated in a larger, 
extensive-igneous processed precursor body 
(HED: howardites, eucrites, and diogenites-assemblage).   
The V-type NEAs \citep{Cruikshank91} remain 
an open question for their origin (from (4)~Vesta?) \citep{Cochran04,Burbine09}.

In this paper we present the orbital and physical properties of Kyoto fireball 
taken by SonotaCo Network, 
including the trajectory, radiant point, geocentric velocity, 
orbit and meteoroid mass (size) 
and further discuss the possible relation to 
the parental binary NEA 2003~YT$_1$ 
by examining its potential dust production mechanisms.

\section{SonotaCo Network}
\label{sonotaco}

The fireball studied here is from the SonotaCo Network database.   
Automated multi-station video observations use more than 100 cameras 
at 27 sites in Japan \citep{So09}\footnote{As of 2018, \url{http://sonotaco.jp/doc/SNM/2018A.txt}}.  
The database is advantaged in the similar type of camera setup of all the network sites.   
The CCD cameras are mostly WATEC series 
with $f$=3.8--12\,mm lens having field of view (FOV) $\approx$ 30$^\circ$ -- 90$^\circ$.  
The video format is digitized in 720 $\times$ 480 or 640 $\times$ 480 pixels AVI  
from the NTSC signal (29.97 frames per second, interlaced), 
and the video field with time resolution of $\simeq$ 0.017\,sec (1/59.94\,sec) is used for measurement.  
Meteors are detected by UFOCaptureHD2 software, and
the data reductions and orbit determinations are conducted
by UFOAnalyzerV2 and UFOOrbitV2 respectively.  
Limiting magnitude for multi-station observations is estimated to be  
apparent magnitude $<$ +3 and absolute magnitude $<$ +2 for each \citep{So09}.

The database includes orbital and physical parameters of meteors, 
such as trajectory (apparent position on the sky plane), radiant point, 
geocentric velocity, orbital elements, brightness (magnitude) 
and height above the sea level\footnote{\url{http://sonotaco.jp/doc/SNM/index.html}}. 
Astrometry and photometric calibrations for meteors are conducted using field stars 
in the background and SKY2000 Master Catalog, Version~4 \citep{Myers01} installed in the UFOAnalyzerV2.   
Single-station observation has some uncertainties of measurements but negligibly small, 
as estimated by position in the sky plane $\sim$0.03$^{\circ}$ \citep{So09}, 
distance to meteor $\lesssim$ 200\,m and elevation angle $\sim$0.02$^\circ$--0.03$^\circ$.
Lens distortion is corrected by background stars' positions fitted by polynomial equation. 
The aperture radius used for the stars is 5pix in the image ($\sim$0.5$^{\circ}$) 
and the sky background is determined within a concentric annulus having projected 
inner and outer radii of 5pix and 7.5pix ($\approx$0.5$^{\circ}$$\sim$0.7$^{\circ}$), respectively. 
For meteors on the other hand, the aperture sets a minimum rectangle 
that covers the total brightness of meteor including its tail,  
and the sky background was subtracted by the field prior to the meteor appearance.  
More than 5 background stars are used to count 
the flux of meteor.  
Then we obtain apparent magnitude of meteor, $m({\rm obs})$. 
The photometric uncertainty (mag) is estimated from
typical uncertainty of comparison stars $\sim$0.5\,mag  
and correction for saturated apparent magnitude of meteor,  
expressed as
\begin{equation}
\sqrt{0.5^2 + \left(m({\rm obs})'  - m({\rm obs}) \right)^2},   
\label{error}
\end{equation}
where $m({\rm obs})'$ is corrected apparent magnitude.  
The $m({\rm obs})'$ is derived from $m({\rm obs})' = m({\rm obs})+ k\,(m({\rm obs}))^2$, 
where $m({\rm obs})$ $< $ 0 and $k$ = $-$0.03.  
Details of the analysis procedure is described in the UFOAnalyzerV2 manual\footnote{\url{
http://sonotaco.com/soft/download/UA2Manual_EN.pdf}}, and private communication with SonotaCo.

\section{Results}
\label{results}

\begin{figure*}[htbp]
\epsscale{1} \plotone{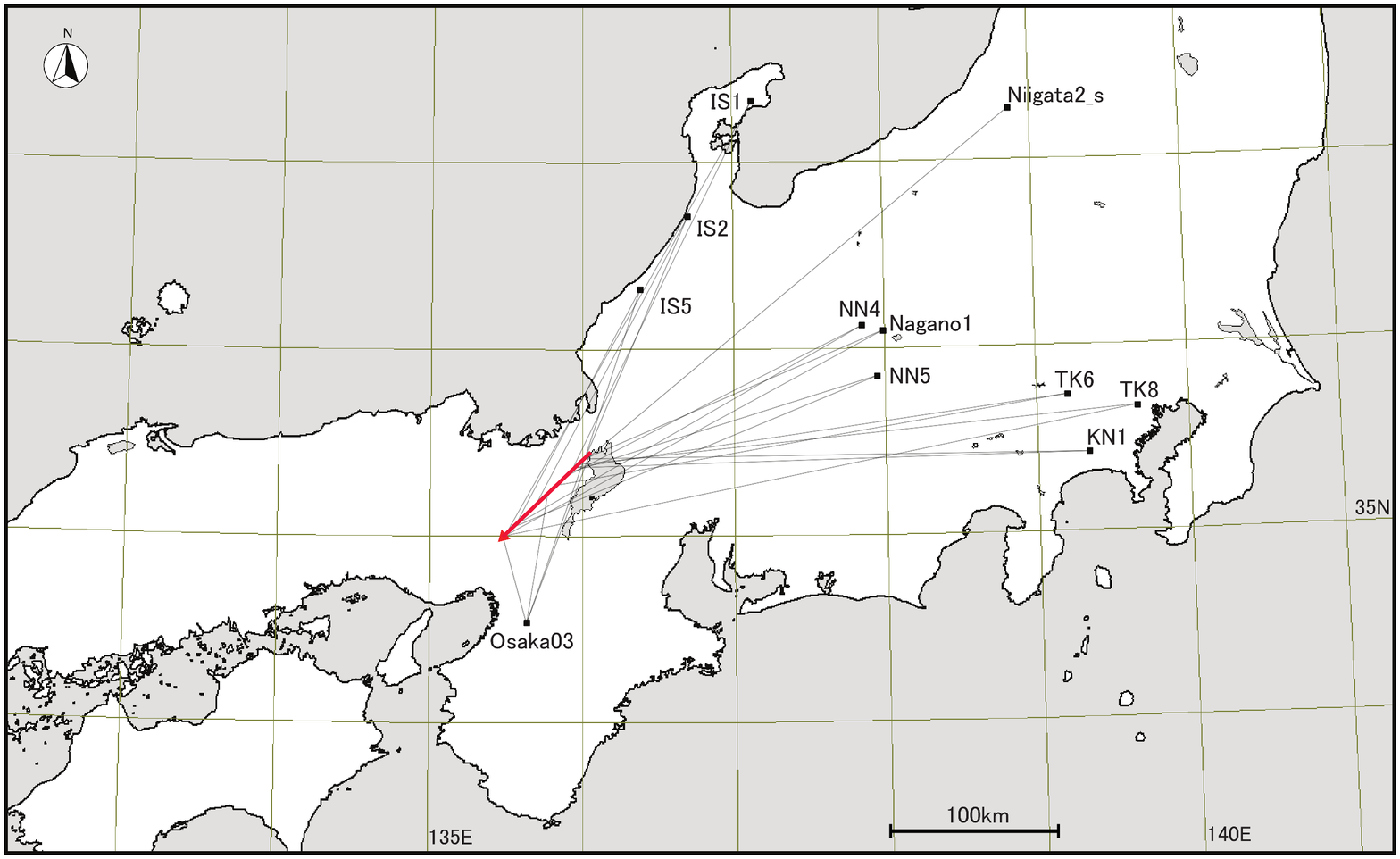} 
\caption{Map showing the projection of fireball atmospheric trajectory (red arrow), 
including eleven observation sites (ID) and the lines of sight (thin-line).  
The direct distance of trajectory is approximately 70\,km. 
The ID is listed in \url{http://sonotaco.jp/doc/SNM/2017C.txt}. 
$\copyright$SonotaCo. 
\label{map}}
\end{figure*}

\begin{figure*}
\gridline{
\fig{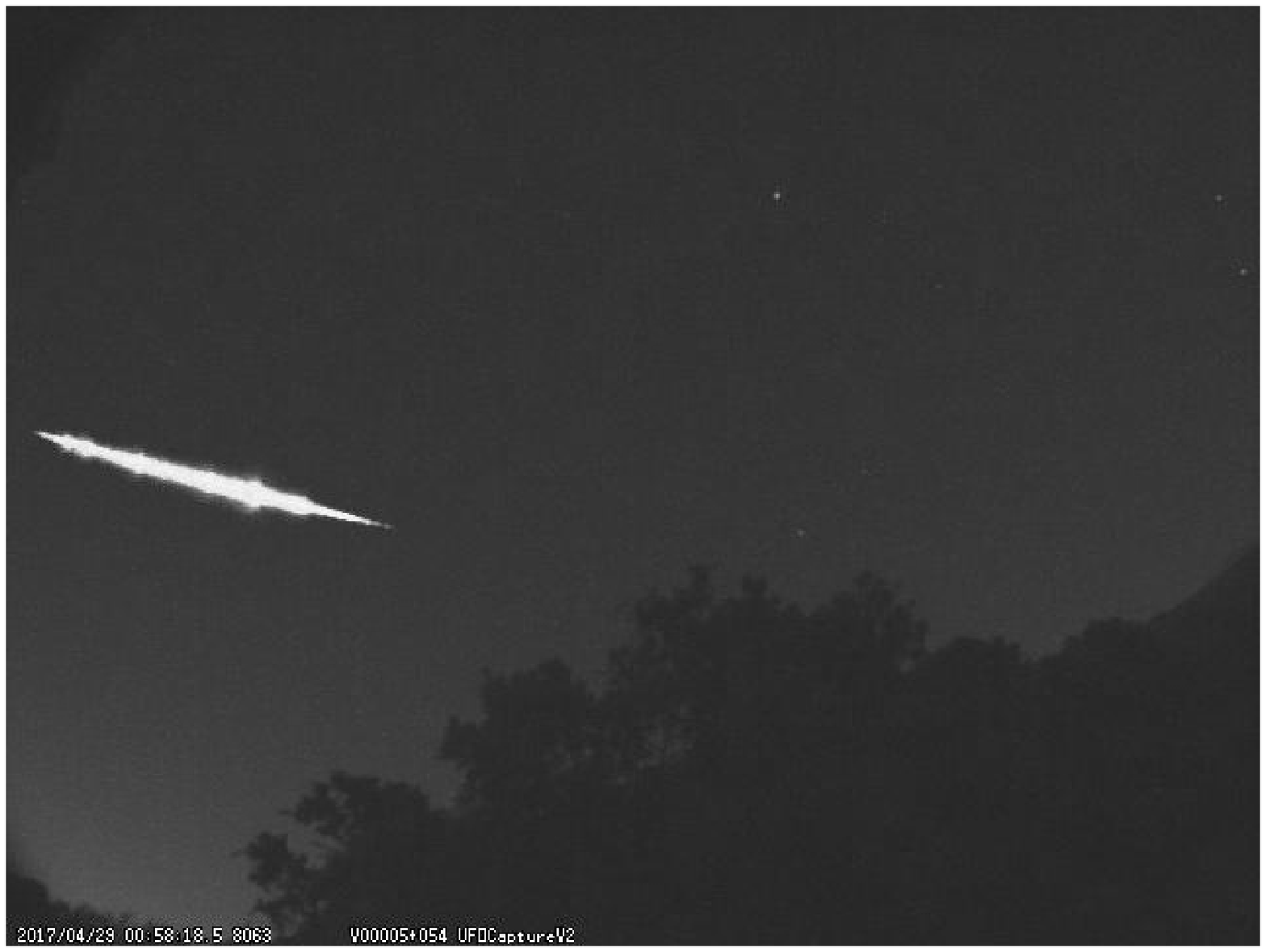}{0.4\textwidth}
{(a) Composite image, with 204 fields (3.40 sec): 
Watec (WAT-231S2), $f$=3\,mm, F0.95 and FOV= $94^{\circ}\times$68$^{\circ}$ 
at Osaka~(Osaka03\_3N).} 
\fig{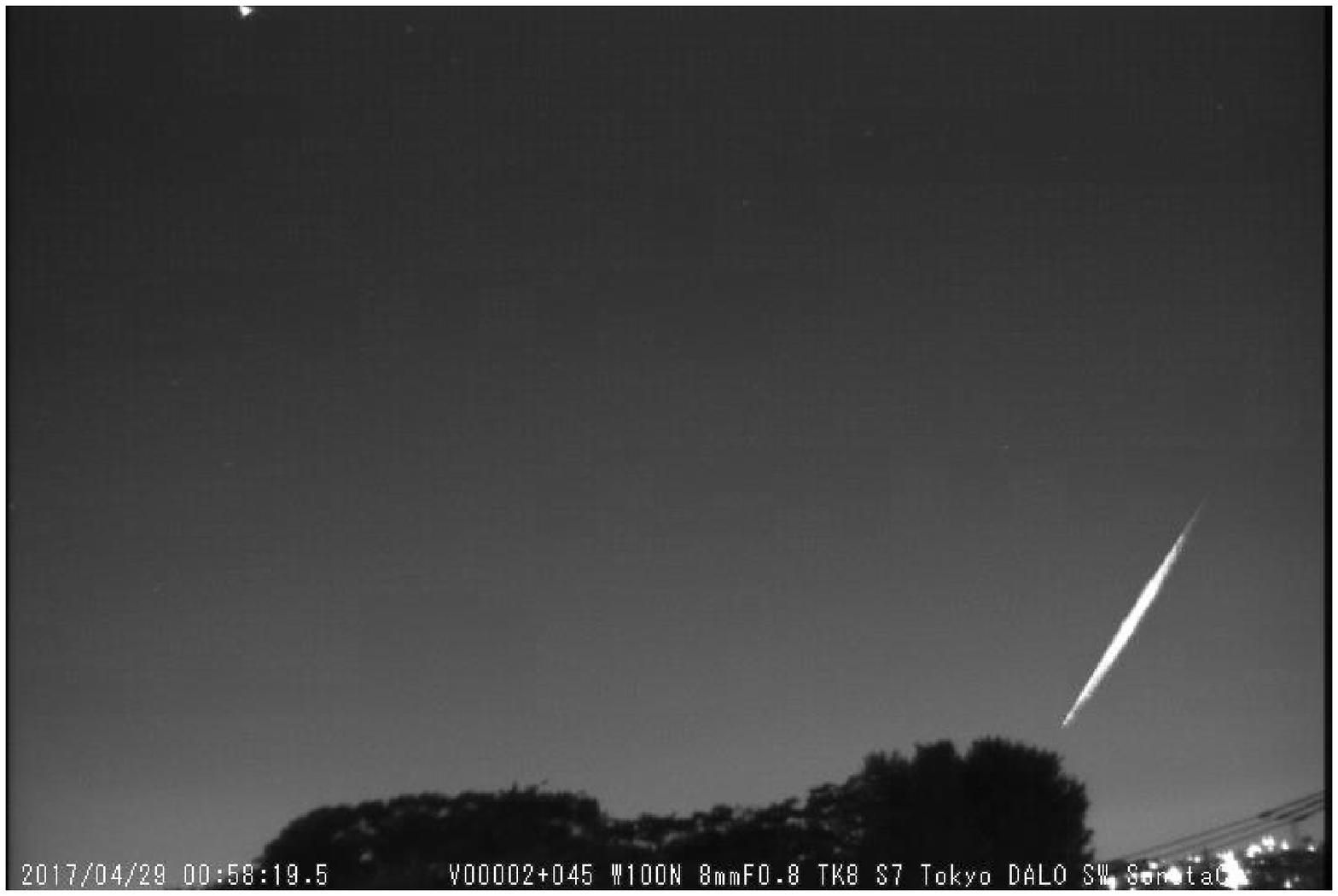}{0.45\textwidth}
{(b) The 173 fields (2.89\,sec): Watec (WAT-100N), $f$=8\,mm, F0.8 and FOV= $45^{\circ}\times$34$^{\circ}$ at Tokyo~(TK8\_S7).}
          }
\gridline{
\fig{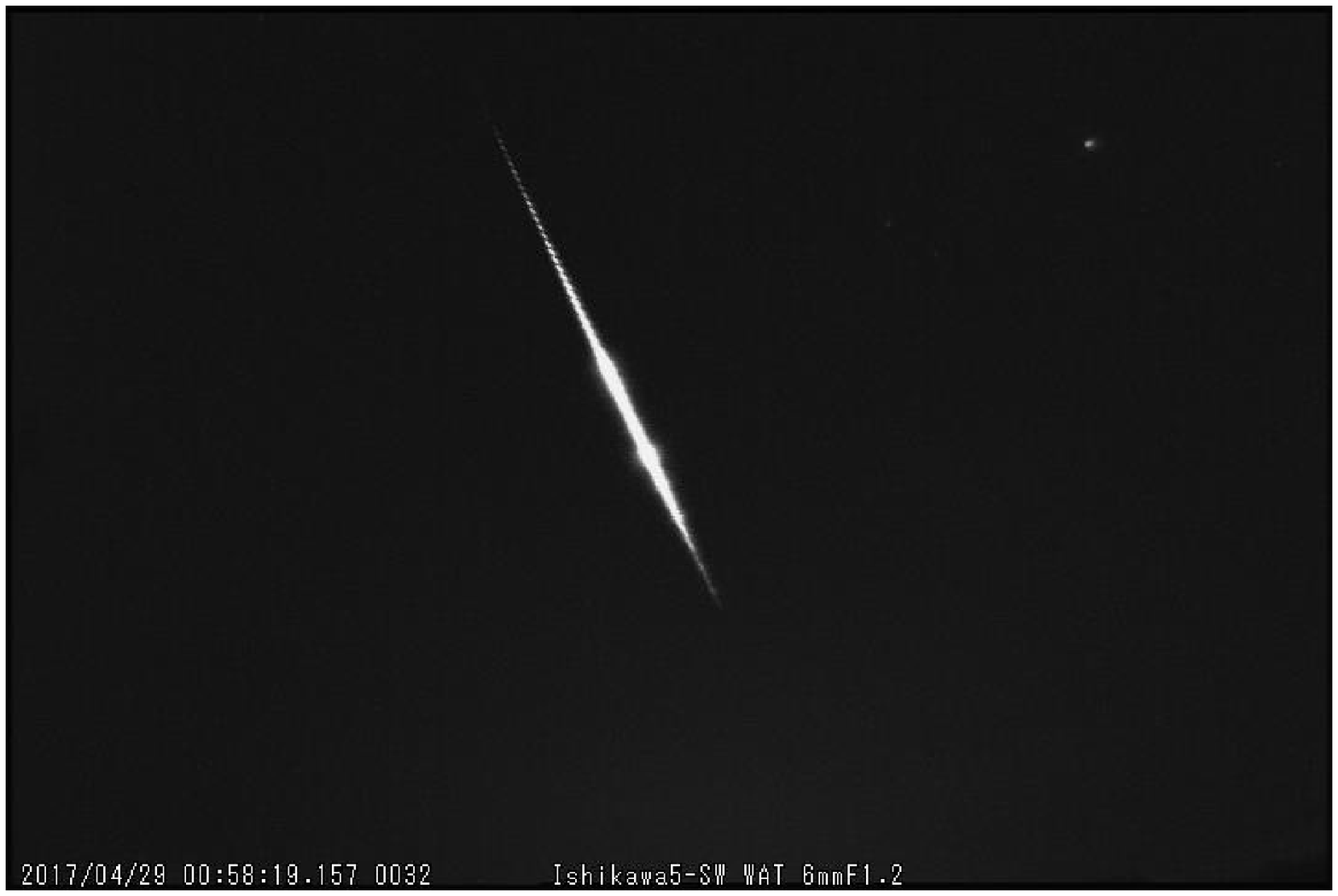}{0.45\textwidth}
{(c) The 164 fields (2.74\,sec): Watec, $f$=6\,mm, F1.2 and FOV= $57^{\circ}\times$43$^{\circ}$ at Ishikawa~(IS5\_SW).}
\fig{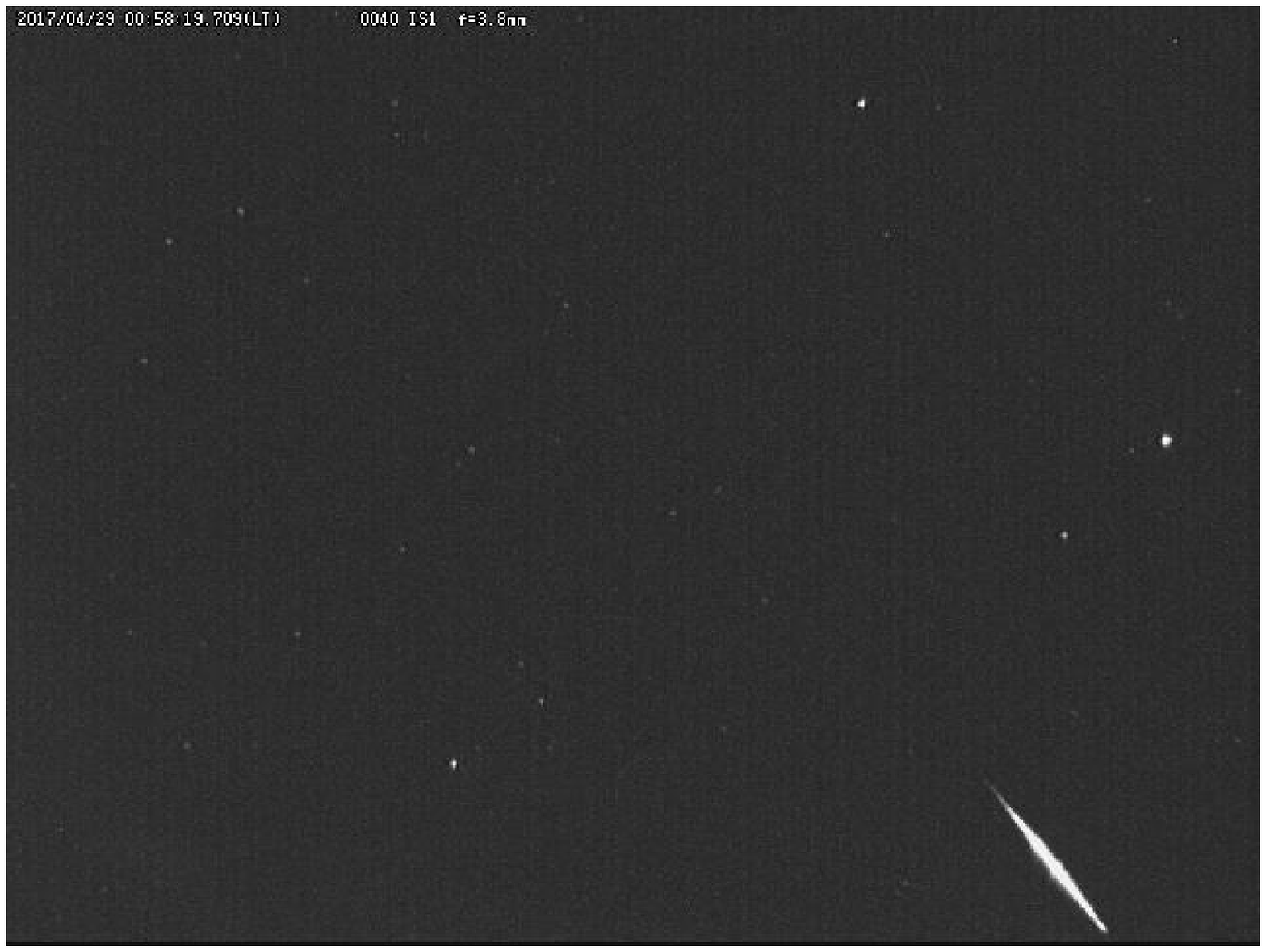}{0.4\textwidth}
{(d) The 162 fields (2.70\,sec): Watec (WAT-100N), $f$=3.8\,mm, F0.8 and FOV= $89^{\circ}\times$69$^{\circ}$ at Ishikawa~(IS1).}}
\gridline{
\fig{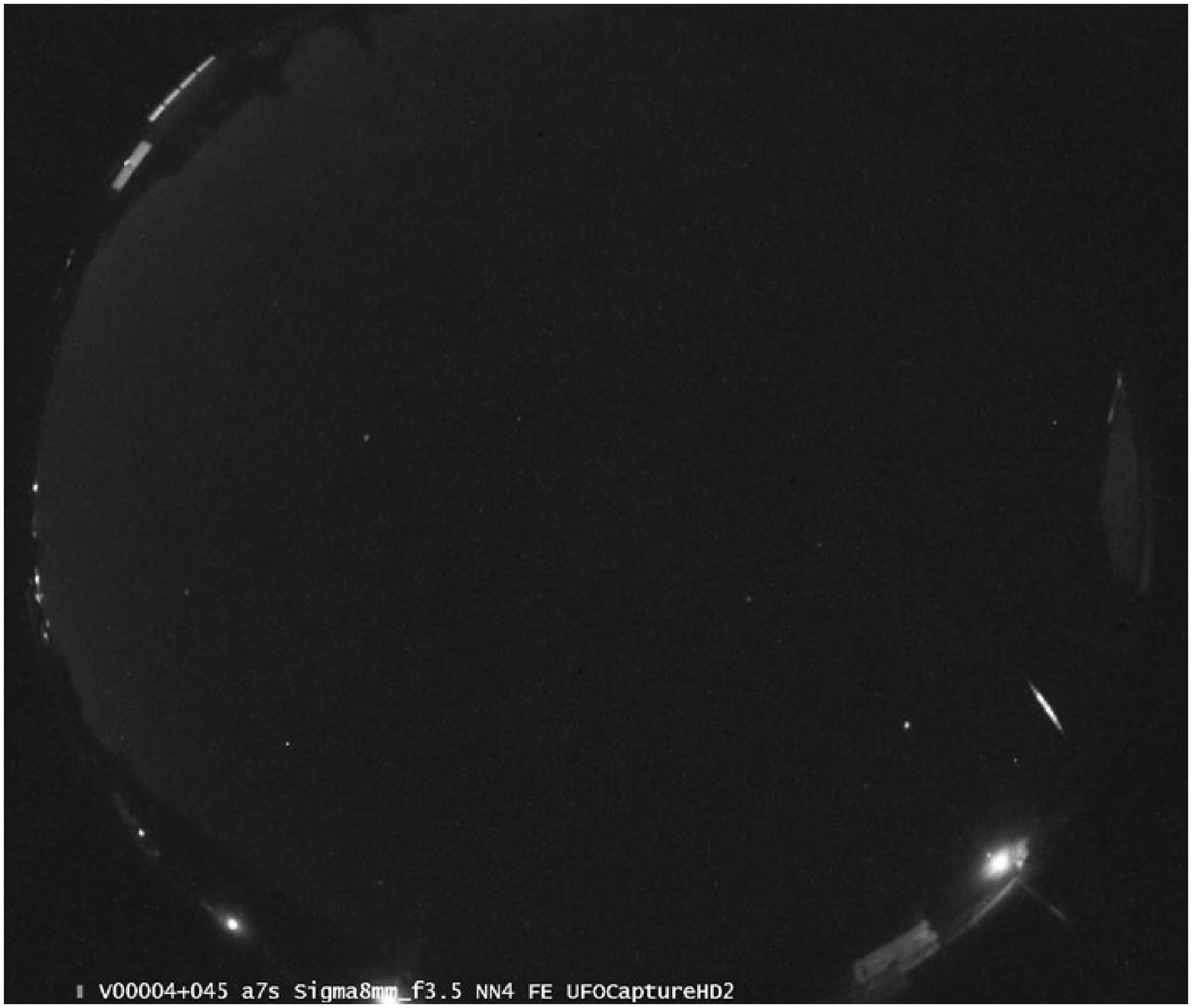}{0.40\textwidth}
{(e) The 104 fields (1.74\,sec): Sony $\alpha$7s (High-definition 1920$\times$1080\,pixel, 60i, interlaced), $f$=8\,mm, F3.5 and fish-eye lens at Nagano~(NN4\_FE).}
\fig{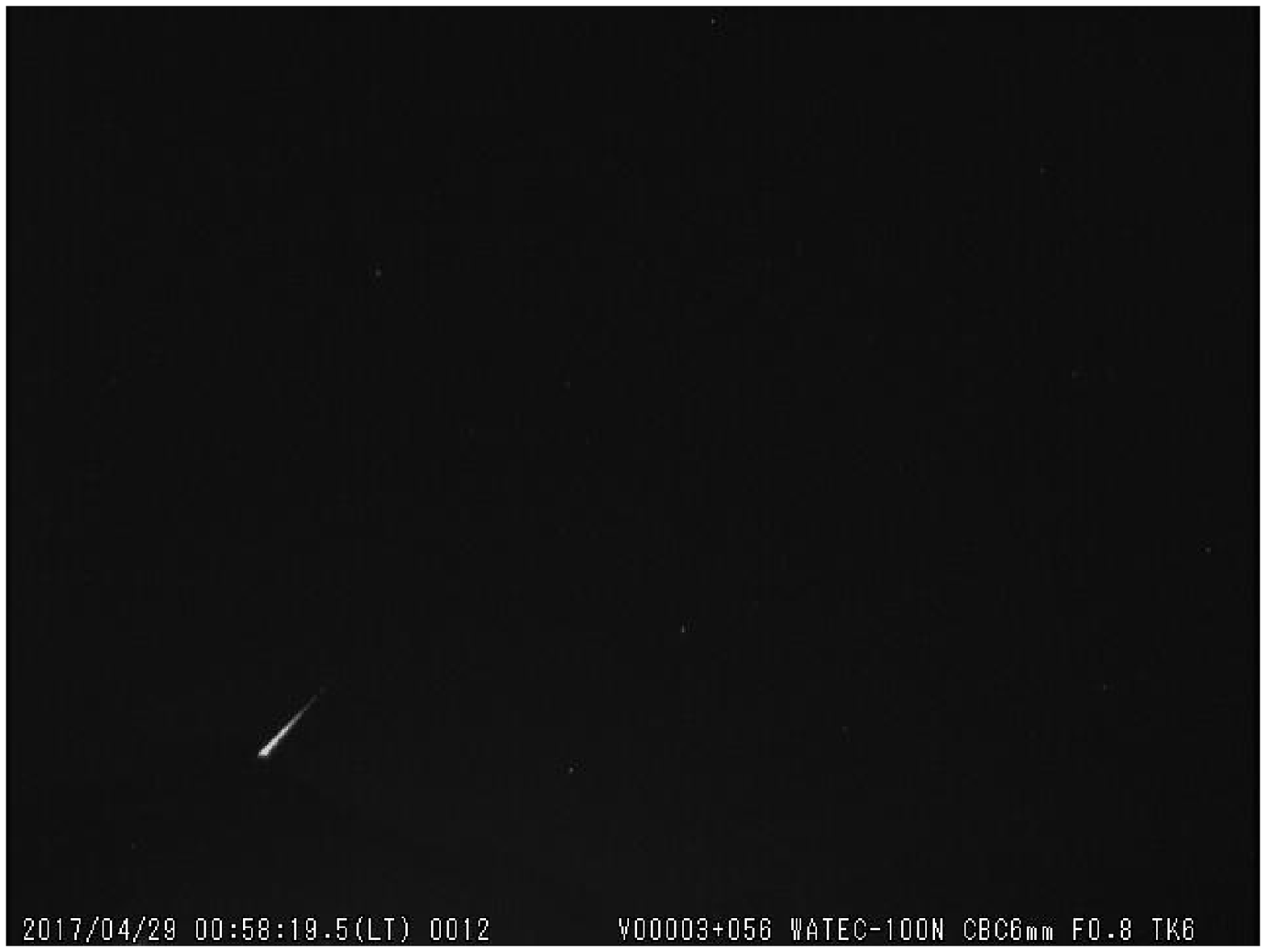}
{0.45\textwidth}{(f) The 41 fields (0.68\,sec): Watec (WAT-100N), $f$=6\,mm, F0.8 and FOV= $57^{\circ}\times$43$^{\circ}$ at Tokyo~(TK6\_w).}
          }
\caption{
Composite images of the fireball recored on UT 2017 April 28 at ${\rm 15^{h}\,58^{m}\,19^{s}}$.
Number of detected video fields (duration time): camera, focal length ($f$), F-number and FOV at observation site (ID).   
The date and time within image is JST.
\label{image}}
\end{figure*}

Fireball trajectory and observing sites 
(ID\footnote{Here we note in the ID that is a assignment for observing site.  
Take TK8\_S7 (Tokyo data) for example, 
the location is expressed as TK8 and the underscore S7 is named after camera.}) 
are shown in Figure~\ref{map}.  
The images of fireball are represented  in Figure~\ref{image}.   
This event was simultaneously detected at eleven sites with twelve cameras.  
The data sets taken at Tokyo (TK8\_S7) and Osaka (Osaka03\_3N) 
have imaged the most part of trajectory, from the beginning to the end.  
Numbers of video fields which have acquired 
the fireball position and brightness are 
159 out of 173 in the Tokyo data 
and 194 out of 204 in the Osaka data respectively.  
Therefore these two data sets are primarily used  
for orbit determinations and photometric measurements.   
Orbital results are listed in Tables~\ref{tabletrajectory}, \ref{radv}, and \ref{orbit}.  
Photometric results are given in Tables~\ref{Tokyo} and \ref{Osaka}.


\subsection{$D$-criterions}
\label{D}

We searched dynamical similarities between the fireball and asteroids 
using distances defined in the orbital elements space, $D$-criterions, 
by comparing $a$ (semimajor axis), $e$ (eccentricity), $i$ (inclination), 
$q$ (perihelion distance), $\omega$ (argument of perihelion)
and $\Omega$ (longitude of ascending node) \citep{Williams19}. 
Three types of $D$-criterions are used to reduce biases therein.  
The first one is $D_{\rm SH}$ \citep{SouthHawkins63} depending mostly on $q$, 
the second is $D'$ \citep{Drummond81} depending mostly on $e$, 
and the third is $D_{\rm ACS}$ \citep{Asher93} neutralizing 
rapid evolutions of the $\omega$ and $\Omega$ with time \cite[cf.][]{Dumitru17}.  
A smaller $D$ indicates closer degree of orbital similarity between two bodies. 
By comparing with the orbit of 2003~YT$_1$ (see Table~\ref{orbit}), 
we find more than one order of magnitude smaller values than 
the significant empirical threshold 
\citep[e.g. $D_{\rm SH}$ $\lesssim$ 0.10--0.20,][]{Williams19}.  
The close-knit orbit interprets that 2003~YT$_1$ is a possible parent body. 
Results are shown in Table~\ref{Dcri}.

We further searched other probable meteors having the similar orbits 
from the SonotaCo data sets in 2007--2018 and 
the European video Meteor Network Database (EDMOND\footnote{\url{http://www.daa.fmph.uniba.sk/edmond}}) 
\citep{Korno14pim3,Korno14me} in 2001-2016, 
but found few compelling cases (Appendix~\ref{SonoEd}).

\subsection{Meteoroid Mass}
\label{memass}

For initial meteoroid mass, 
the classical meteor luminous model \citep[][]{Bronshten83,CBE98} 
has been used but with non-negligible uncertainty in ablation coefficient.    
Instead, we have made a new meteor luminous model  
as described in the Appendix~\ref{appendixA}. 

\begin{figure*}[htbp]
\epsscale{1} 
\plotone{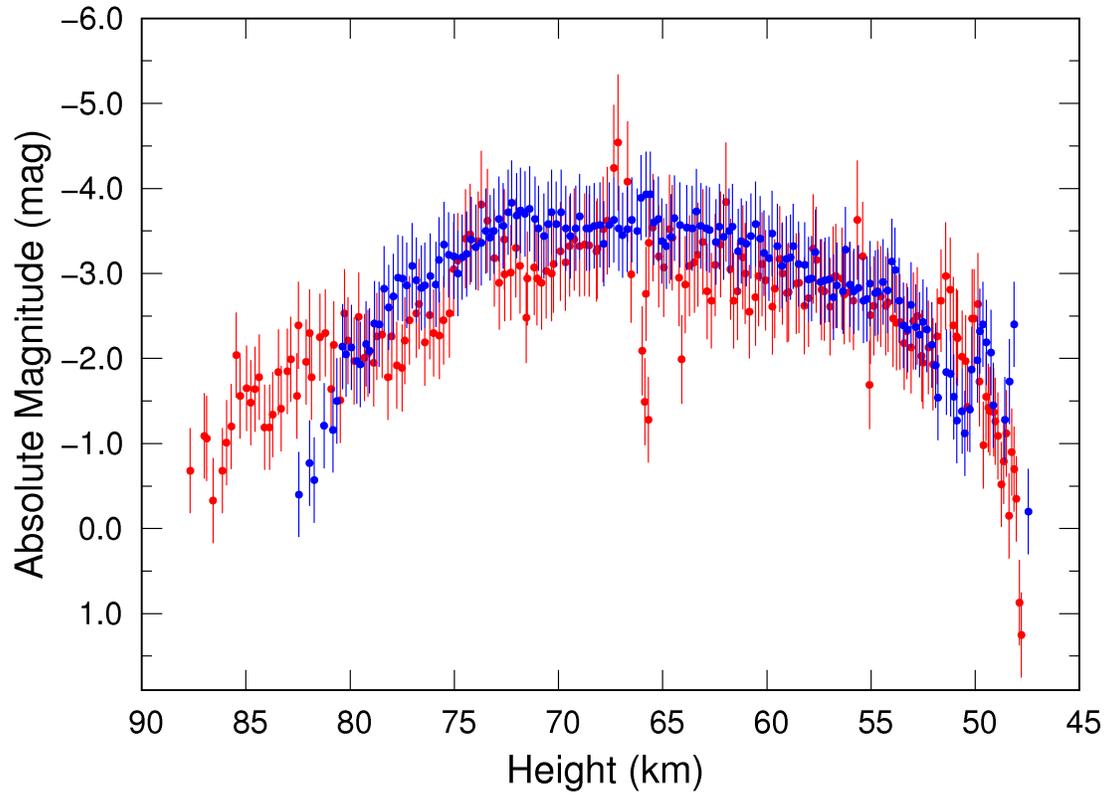} 
\caption{Light curves of fireball measured at Tokyo (Blue) and Osaka (Red).  
Absolute magnitudes as a function of height are plotted from Tables~\ref{Tokyo} and \ref{Osaka}. 
The weighted mean of maximum brightness is $-$4.10$\pm$0.42\,mag.  
The uncertainty of height is within the circle.
}
\label{lightcurve}
\end{figure*}

The total mass of meteoroid (source of fireball), $m$ (g), 
can be estimated from the light curves (Figure~\ref{lightcurve}) 
using the new luminous model (Equation~(\ref{newmodel})),  
given by 
\begin{equation}
m =  \sum_{\rm N} \left[ \frac{2\,I}{\tau v^2} \left(\frac{2}{\sigma v^2} - 1  \right)^{-1} \int_0^{t} dt \right]_{N}, 
\label{intmag}
\end{equation}
where 
$N$ is the number sign of video field (see Tables~\ref{Tokyo} and \ref{Osaka}), 
$I$ is the meteor luminosity, 
$\tau$ is the luminous efficiency, 
$v$ is the meteor velocity (cm\,s$^{-1}$),  
$\sigma$ is the ablation coefficient (s$^{2}$\,km$^{-2}$) 
and $t$ is the time (s).
We define the meteor luminosity in visual magnitude-based units 
as $I = 10^{-0.4 M_{\rm v}}$, where $M_{\rm v}$ is the absolute 
magnitude (as seen from distance of 100\,km).

The luminous efficiency, $\tau$, is the fraction of 
a meteoroid's instantaneous kinetic energy loss 
converted into light in a particular band-pass.  
The uncertainty within is substantial (0.05$\sim$10\,s\,\%)
as it depends on many factors, e.g. 
the speed, mass, composition of meteoroid and the height 
at which it ablates (different flow regimes)  
and spectral sensitivity of detector 
\citep[cf.][]{Weryk13pss,Subasinghe18} \citep[See the revew,][]{Popova19}. 
For this study, we use the velocity dependence \citep[Table~1 in][]{CeplechaMcCrosky76} 
considering the performance of CCD cameras (e.g. low resolution).   
Setting $v$=23.7\,km\,s$^{-1}$ 
finds $\tau$=5$\times$10$^{-13}$\,erg$^{-1}$\,s\,0mag.  
The $\tau$-value corresponds to 0.75\% efficiency.  
The conversion is given by multiplying 
1.5$\times$10$^{10}$\,erg\,s$^{-1}$\,0mag$^{-1}$, 
i.e. the luminous energy equivalent to 
zero magnitude in visual \cite[Table~VI in][]{CBE98}.

The critical bulk density, $\rho$, for the meteoroid and 2003~YT$_1$ is estimated.  
An asteroid shape is approximated as an ellipsoid 
with axes a $\ge$ b = c, in rotation about the c-axis.  
A limit to the ratio of the equatorial axes is, 
$f$=a/b=10$^{0.4 \Delta m}$, where $\Delta m$ is the light curve amplitude.  
Rotation around the c-axis with period, $P_{\rm rot}$, 
gives a condition that 
the gravitational acceleration 
is greater than 
the centripetal acceleration which is the largest at the top of the shape.  
The net acceleration toward the center of a rotating object is $>$ 0, 
giving the relation as \citep[Equation (4) of][]{JL10},  
\begin{equation}
\rho > \left(\frac{3 \pi}{G P_{\rm rot}^2}\right) \left(\frac{\rm a}{\rm b}\right)^2, 
\label{rho}
\end{equation}
where $G$ is the gravitational constant.
We substitute 
$G$ = 6.67 $\times$ 10$^{-11}$m$^3$\,kg$^{-1}$\,s$^{-2}$, 
$P_{\rm rot}$ = 2.343\,hr, 
$\Delta m$ = 0.16\,mag (i.e. $f$=a/b=1.16) 
into Equation~(\ref{rho}), 
then obtain $\rho$ $\gtrsim$ 2700\,kg\,m$^{-3}$.  
This is consistent with the lower limit for 
rubble pile asteroids with diameters of 0.3--10\,km ($\rho$ = 2.7\,g\,cm$^{-3}$), 
as formulated by the observed light curve amplitude versus spin rate \citep{Pravec05SASS}.
The proposed bulk density 2010$\pm$700\,kg\,m$^{-3}$ \citep{Brooks06} 
may be uncertain due to the assumption of circular orbit of the secondary.  
The orbit is actually eccentric \citep[$e$ $\simeq$ 0.18,][]{FangMargot12a} \citep[cf.][]{Pravec16}.

Substituting $v$(=$v_{\rm g}$)=2.37$\times$10$^{6}$\,cm\,s$^{-1}$, 
$\tau$=0.75\%, 
$\sigma$ = 0.0017\,s$^{2}$\,km$^{-2}$ (Appendix~\ref{appendixA}) 
and 
$t$ = 0.017\,s 
into Equation~(\ref{intmag}), 
we obtain $m$ (see Tables~\ref{Tokyo} and \ref{Osaka}).  
The weighted mean of total mass is $m$=29$\pm$1\,g, 
corresponding to meteoroid size $a_{\rm s}$=2.7$\pm$0.1\,cm for $\rho$ = 2700\,kg\,m$^{-3}$.  
For reference, the classical luminous model (Equation~(\ref{I_C98})) is applied too. 
The resulting masses are compared in Table~{\ref{modelmass}}.

\section{Discussion}

Here, we recapitulate the binary formation process of 2003~YT$_1$ 
and evaluate possible dust production mechanisms 
for mm-cm scale particles.

\subsection{Binary Formation} 
\label{YORP}

The 2003~YT$_1$ binary system is presume to be formed from 
a breakup/fission by rotational instability with YORP spin-up.  
The primary with $D_{\rm p}$ $\lesssim$ 10\,km and 
the normalized total angular momentum of the binary system $\alpha_{\rm L}$
\footnote{
The $\alpha_{\rm L}$ is the ratio of the total angular momentum of the system 
to the angular momentum of a critically spinning spherical body. 
The spherical body is comprised of the mass and volume equivalent to the two objects of the binary system.  
The internal friction angle is 90$^\circ$ \citep{Pravec07Icar}.}  
=1.13 
suggest that the 2003~YT$_1$ binary system was formed from a precursor body spinning 
at the critical rate, resulting in fission and mass shedding 
\citep[Group A in Table~1 and Figures~2 and 3 from][]{Pravec07Icar} \citep[Reviewed in][]{Margot15aste,Walsh15aste}.  
The 2003~YT$_1$ primary rotates ($P_{\rm rot}$=2.343\,hr) closely
to the spin barrier period $\sim$ 2.2\,hr \citep{WHP09,Chang_Ip_15}. 
This can reasonably lead to a rotational breakup 
when centrifugal forces have exceeded 
the gravitational and cohesive forces \citep{Pravec08Icar}.

We calculate the YORP timescale of the spin, $\tau_{\rm Y}$, 
using the ratio of the rotational angular momentum, $L$, to the torque, $T$.  
The relation is given by \cite{JHA15} as,
\begin{equation}
\tau_{\rm Y} \sim K\,D_e^2\,R_h^2, 
\label{tauY}
\end{equation}
where $K$ is a constant, 
$D_{\rm e}$ is the asteroid diameter (km) 
and $R_{\rm h}$ is the heliocentric distance (AU).  
The value of constant $K$ is experimentally estimated 
from published measurements of YORP acceleration 
in seven well-characterized asteroids \citep[Table~2 from][]{RozitisGreen13}.
Scaling $K$ to the bulk density of primary $\rho$ = 2700\,kg\,m$^{-3}$ 
and its rotation period $P_{\rm rot}$=2.343\,hr, 
we find $K \sim$ 5$\times$10$^{13}$\,s\,km$^{-2}$\,AU$^{-2}$.  
By Equation~(\ref{tauY}), 
the primary with $D_{\rm e}$(=$D_{\rm p}$)=1.1\,km 
orbiting at $R_h \sim$1.11\,AU 
takes $\tau_{\rm Y} \sim$ 2\,Myr.   
This is consistent with the previous study \citep[$\sim$1\,Myr,][]{Vokrouhlick15} 
and much shorter than 
the catastrophic collisional lifetime for 1-km NEAs \citep[$\sim$100\,Myr,][]{Bottke94hdtc} 
(see also Section \ref{impact}).   
The YORP spin-up plays a contributory role.

The cohesive strength is a required parameter for asteroids rotating 
near or faster than the spin-barrier to resist rotational forces \citep[][]{Scheeres10}.   
The strength at a rotational breakup of a body is 
estimated by 
$S_{\rm c} \sim \rho \left(D_{\rm s} / D_{\rm p} \right)(\Delta v)^2$ \cite[Equation (5) of][]{JHA15}, 
where $D_{\rm p}$ and $D_{\rm s}$ are 
the dispersed fragmental sizes of primary and secondary respectively, 
$\Delta v$ is the excess velocity of escaping fragments, 
assumed comparable to the escape velocity ($v_{\rm e}$) from the primary, 
and $\rho$ is the bulk density.  
With the same value for $\rho$ (see section~\ref{memass}) and substituting 
($D_{\rm s}$/$D_{\rm p}$) = 0.19 (the diameter ratio of the secondary to primary), 
and $\Delta v$ (=$v_{\rm e}$) = 0.68~m\,s$^{-1}$, we find $S_{\rm c} \sim$ 240~N\,m$^{-2}$.  
This value is comparable to weak, van~der~Waals forces ($\sim$10--100~N\,m$^{-2}$) 
bounded in a modeled rubble-pile asteroid \citep[][]{Scheeres18}, 
while 10$^5$ $\times$ weaker than those of competent rocks (10$^7$--10$^8$ N\,m$^{-2}$).  
Therefore, given a rubble-pile structure,  
a rotational breakup/fission is a probable process for the 2003~YT$_1$ binary formation.

The breakup/fission period is inferred from 
the spin asynchronous state of 2003~YT$_1$ binary system in the present day.  
The timescale from asynchronous to synchronous state, $\tau_{\rm sync}$, 
limits to the age of the binary system.  
Two models are applied for 2003~YT$_1$ using the data of 
known synchronous binary \citep[Table~3 in][]{FangMargot12a}. 
One estimates $\tau_{\rm sync}$=10$^{7-8}$\,yr 
by the tidal Love number proportional to the radius \citep {Goldreich09}, 
another estimates $\tau_{\rm sync}$=10$^{4-5}$\,yr 
by the tidal Love number inversely proportional to the radius \citep{Jacobson11}.  
The former just agrees with the large-sized binaries having primary with $D_{\rm e} \sim$ 4\,km,   
on the other hand, the latter fits well for smaller-sized objects too (down to $D_{\rm e} \sim$ 0.4\,km).  
For 2003~YT$_1$ we thus take $\tau_{\rm sync}$=10$^{4-5}$ yr \citep{FangMargot12a}. 
The interpretation is that this binary is age of $<$ 10$^{4}$\,yr, 
comparable with the upper limit to 
meteoroid stream lifetime $<$ 10$^{4}$\,yr \citep{PJL05}.    

Another example is proposed by the small-sized V-type NEA pair ($D_{\rm e}$$\sim$25--50\,m) 
also having the young age of separation $<$ 10$^{4}$\,yr 
($D^{'}$=0.0035 for 2017~SN$_{16}$ and 2018~RY$_{7}$) \citep{Moskovitz19}.      
The YORP-driven breakups for the (sub)km-sized bodies 
may suggest moderately recent events.

\subsection{Dust Production Mechanisms}

We look into possible dust production mechanisms from 2003~YT$_1$.  
The consequences of YORP-driven breakups 
are reported from the (sub)km-sized main-belt asteroids, 
as exampled by 
P/2010~A2 \citep{Jewitt10Nat,Jewitt13ApJ,Agarwal13ApJ},  
P/2013~R3 \citep{Jewitt14R3,Jewitt17R3,Hirabayashi14ApJ}
and (6478) Gault \citep{Ye19,Moreno19,Kleyna19,Jewitt19,Chandler19,Hui19}.  
Additionally, other different mechanisms may work together, 
e.g. impact for P/2010~A2 
and 
outgassing torques from sublimated ice for P/2013~R3 \citep[][]{JHA15}. 
Here, we estimate 
breakup/fission (rotational instability), resurfacing, impact, 
thermal fracture, photoionization, radiation pressure sweeping
and sublimation of water ice.

\subsubsection{Breakup/Fission (Rotational Instability)}
\label{break}

Binary NEAs show a trend to have 
the large values of thermal inertia $\Gamma \gtrsim$\,400\,J\,m$^{-2}$\,s$^{-0.5}$\,K$^{-1}$, 
typically twice those of non-binary NEAs, 
suggesting the fine regoliths were swept away 
during the YORP-induced binary formation \citep{Walsh08,Delbo11}.  
For 2003~YT$_1$,  
it would be difficult to determine 
the sizes and speeds of released dust particles at the presumed breakup time, 
whereas the measured values of the recent precedents infer 
the large particles (mm--cm scale) with nearly the gravitational escape speeds $\lesssim$\,1\,m\,s$^{-1}$ 
\citep[cf. $a_{\rm s}$=6\,mm--40\,cm from P/2010~A2, $\sim$1\,cm from P/2013~R3 and $\lesssim$ 
1\,cm from Gault, ][]{Jewitt13ApJ,Jewitt14R3,Jewitt19}.  
On the process, resurfacing could be partly involved \citep[Gault,][]{Marsset19}. 
The similar situation might be expected for 2003~YT$_1$.   
The dust particles are, if released, supposed to reach 
the Earth within the typical stream lifetime (10$^{4}$\,yr).  
The short distance to the Earth orbit, 
e.g. $\Delta r$=0.0026--0.0279\,AU (Table~\ref{radv}), 
may help.  
Accordingly, the rotational breakup/fission ejecting 
the mm-cm scale dust particles 
is considered as a likely cause.

\subsubsection{Resurfacing}
\label{resurface}

Planetary encounters, space weathering and thermal processes 
could induce resurfacing, which 
might lose dust particles on the surfaces to some extent.  
For example, the timescale for  
Q-type NEAs to be refreshed into S-type (at 1\,AU and $q \lesssim$0.9\,AU) 
is estimated 10$^{5-7}$\,yr 
by planetary encounters \citep{Nesvorn10,Binzel10}, 
space weathering \citep{Graves18} and thermal processes \citep{Graves19}.  
For V-type NEAs, the aftermaths of those processes are unclear \citep[space weathering, ][]{Pieters12,Fulvio16}, 
while the timescale of resurfacing itself 
seems to be 10--1000 times longer than the typical stream lifetime.  
The resurfacing is thus unlikely to be responsible for 
releasing the source of meteors.

\subsubsection{Impact}
\label{impact}

Impacts can cause catastrophic disruption of asteroids and/or dust production.    
The catastrophic disruption is defined as 
the impact resulting in losing a half of the target's mass.  
The specific impact energy threshold is expressed as 
$Q_{\rm D}^{*}$ = (1/2)($D_{\rm i}$/$D_{\rm t}$)$^3$$\Delta V_{\rm NEA}^2$, 
where $D_{\rm i}$, $D_{\rm t}$ and $\Delta V_{\rm NEA}$ are 
the size of the impactor and the target (an assumed precursor body)
and the relative velocity among NEAs, respectively \citep[][]{Jutzi10}.   
With $Q_{\rm D}^{*}$$\sim$1400\,J\,kg$^{-1}$
for catastrophic disruptions of stone meteorites \citep{Flynn04,Flynn18}, 
$D_{t}$$\approx$$D_{\rm p}$=1100\,m (assuming the primary size occupying $>$ 80\% of the precursor body) 
and $\Delta V_{\rm NEA}$ = 17--20\,km\,s$^{-1}$\citep[][]{Bottke94hdtc,Jeffers01},  
we find $D_{\rm i}$ $\sim$ 20\,m. 
This catastrophic event is inferred from  
the interval between impacts, $\tau_{\rm col}$ \citep{Davis2002}, 
\begin{equation}
\tau_{\rm col} \simeq \frac{4}{\pi (D_{\rm t}+D_{\rm i})^2\,P_{\rm NEA}\,N_{\rm i}(\geqslant D_{\rm i})},
\label{tau}
\end{equation}
where $P_{\rm NEA}$ is the collision frequency per unit area in the near-Earth region (km$^{-2}$\,yr$^{-1}$), 
and $N_{\rm i}\,(\geqslant\,D_{\rm i})$ is the cumulative number of impactor larger than $D_{\rm i}$.  
The NEA cumulative size distribution is measured by {\it WISE/NEOWISE},   
$N_{\rm i}$$(D_{\rm i}$$\geqslant$140\,${\rm m}$) $\simeq$ 
13200 $\times$ $\left(140\,{\rm m}/D_{\rm i}\right)^{1.32}$ \citep{Mainzer11}, 
and we presumably extend the equation down to 20\,m in diameters.    
With $D_{\rm i}$ $\sim$ 20\,m, 
$N_{\rm i}$($\gtrsim$ 20\,m) $\sim$ 1.7$\times$10$^{5}$,
$D_{\rm t}$$\approx$$D_{\rm p}$=1.1\,km (see above) 
and 
$P_{\rm NEA}$=1.5 $\times$ $10^{-17}$ km$^{-2}$\,yr$^{-1}$\citep{Bottke94hdtc}, 
we find $\tau_{\rm col} \gtrsim $ $10^{11}$\,yr. 
This is much longer than $\tau_{\rm sync}$=10$^{4-5}$\,yr \citep{FangMargot12a} 
and the mean dynamical lifetime of NEAs $\sim$10$^{6}$\,yr \citep{Bottke02,Morbidelli02aste}, 
suggesting absence of catastrophic event (cf. Section \ref{YORP}).

On the other hand, micrometeorite impacts may result in ejecting the meteoroid-sized particles.  
The velocity distribution for micrometeorites near the Earth, $U$, 
is 12\,$\sim$\,70\,km\,s$^{-1}$ \citep[cf. radar observations, ][]{Nesvorn10,Janches14,Carrillo15}.  
For equal target and impactor densities, the ratio of the ejecta mass, $m_{\rm e}$, 
traveling faster than the escape velocity, $v_{\rm e}$, to impactor mass, $m_{\rm i}$ is related by 
\begin{equation}
m_{\rm e}/m_{\rm i} = A(v_{\rm e}/U)^\alpha, 
\label{mratio}
\end{equation}
where $A$ $\sim$ 0.01, $\alpha$ $\sim$ $-$1.5 \citep{HousenHolsa11}.  
Substitution $m_{\rm e}$$\sim$ 30\,g (fireball mass), $v_{\rm e}$=0.68\,m\,s$^{-1}$ 
and $U$=12\,--\,70\,km\,s$^{-1}$ into Equation~(\ref{mratio}), 
we find that micrometeorite impactors in the size range 
0.4\,mm $\leqslant$ $a_{\rm i}$ $\leqslant$ 1\,mm 
($m_{\rm i}$=(0.1--1.3)$\times$10$^{-3}$\,g with $\rho$ = 2.7\,g\,cm$^{-3}$) 
can eject the cm-sized dust particles.  
The perpendicular impact strength $>$ 10$^{11}$\,N\,m$^{-2}$ 
is estimated from the equation of impact force per unit area
given by $m_{\rm i}$\,$U$/$\delta t$ $\times$ 4/$\pi$\,$a_{\rm i}^2$, where 
$\delta t$ = $a_{\rm i}$/$U$ is the assumed extend impact time (s). 
The value is by orders of stronger than the compressive strengths of stone meteorites  
$\sim$10$^{8}$\,N\,m$^{-2}$ \citep[][]{Flynn18}, 
suggesting micrometeorites are certainly smashing the surface.  
In this case, many of unknown relevant physical parameters 
(e.g. impact frequency, population of micrometeorite near 2003~YT$_1$) 
prevent exact estimation, however, offers probable insight for dust production.

\subsubsection{Thermal Fracture}
Thermal fracture and fatigue of the asteroid surfaces 
can be caused by desiccation stress, 
with the release of dust particles \citep[][]{JL10}.  
For 2003~YT$_1$, the peak perihelion temperature $T_{q} \sim$440\,K 
is about half or less of those of near-Sun asteroids \citep[cf. Phaethon,][]{Jewitt13}, 
while the thermal stress $\lesssim$ 50\,MPa is 
somewhat responsible for breakdown of the rocky surfaces of 
most asteroids in the inner Solar System \citep[Figure~9(b) in][]{Molaro15}.   
The characteristic speeds of dust particles produced by 
thermal disintegration can be computed 
by conversion from thermal strain energy into 
kinetic energy of ejected dust particles.  
We use the required conversion efficiency, 
$\eta$, given by \cite[cf. Equation (3) of][]{JL10} 
\begin{equation}
\eta \sim \left( \frac{v_e}{\alpha\,\delta T} \right)^2 \left(\frac{\rho}{Y}\right), 
\label{eta}
\end{equation}
where, again $v_e$ = 0.68\,m\,s$^{-1}$, 
$\alpha$ $\sim$ 10$^{-5}$\,K$^{-1}$ is the characteristic thermal expansivity of rock \citep{Lauriello1974,Richter1974}, 
$\delta T\sim$80\,K is the temperature variation between the $q$ and aphelion,  
and $Y$ = (1--10) $\times$ 10$^{10}$ N\,m$^{-2}$ are 
Young's moduli for rock in general \citep[][]{Pariseau2006}.  
With $\rho$ as above we find $\eta \gtrsim$ 2--20\,\% is needed 
for the velocities of ejected dust particles to surpass the escape velocity.
The value of conversion efficiency is small enough 
for most dust particles to be launched into interplanetary space.  

Note that micron-sized particles are observed  
from the Phaethon tails at perihelion, possibly 
produced by a combination of thermal fracture 
and radiation pressure \citep{Jewitt13ApJ,Hui17}.  
Such tiny particles are distinct from 
the mm-cm scale dust.    
Larger, mass-dominant particles could be 
launched, but the acquisition of more and 
better data for estimation is waited \citep{Jewitt18HST,Jewitt19Thermal}. 
This mechanism is hence pending.

\subsubsection{Photoionization}  

Photoionization by solar UV induces electrostatic forces to eject very small particles. 
For a 1\,km-diameter asteroid 2003~YT$_1$, the critical size is estimated 
to be $a_e \lesssim$ 4 $\micron$ \citep[Equation (12) of ][]{JHA15}. 
Therefore, mm--cm scale particles cannot be launched.  
We conclude this process is improbable.

\subsubsection{Radiation Pressure Sweeping}     
\label{radswe}

Small dust particles on the surface of asteroids, 
if they briefly lose contact forces, 
can be stripped away by radiation pressure sweeping. 
By equating the net surface acceleration (gravitational and centripetal) 
with the acceleration due to radiation pressure, 
we estimate the critical size to be swept away, $a_{\rm rad}$ ($\micron$), 
with Equation (6) of \cite{JL10}
\begin{equation}
a_{\rm rad} \sim \frac{3\,g_\odot}{2\,\pi\,R_{\rm AU}^2\,f^{1/2}\,D_e} \left[\frac{G\,\rho}{f^2} - \frac{3\,\pi}{P_{\rm rot}^2} \right]^{-1}, 
\label{rad}
\end{equation}
where, $g_\odot$ is the gravitational acceleration to the Sun at 1\,AU, 
$R_{\rm AU}$ is the heliocentric distance expressed in AU, 
$f$ is the limit to the axis ratio (=a/b) 
and 
$G$ is the gravitational constant.  
We substitute 
$g_\odot$ = 0.006\,m\,s$^{-2}$, 
$R_{\rm AU}$ = 0.786 (non-dimensional),
and 
adopt the same values of 
$f$, 
$G$, 
$D_{\rm e}$(=$D_{\rm p}$), $\rho$ and $P_{\rm rot}$ 
(as applied so far) into Equation~(\ref{rad}), 
then obtain $a_{\rm rad}$ $\sim$ 2,900 $\micron$ $\approx$ 3\,mm.  
The mm-sized dust particles can be swept 
by radiation pressure from 2003~YT$_1$, 
could be the source of meteors.   
Even if they arrived at the Earth, 
the relatively small size and slow velocity would be  
orders of magnitude too faint meteors $m_{\rm obs}$ $\sim +$5\,mag \citep[Table~I in][]{Lindblad87}
for most optical surveys.   
By contrast, the cm-sized dust particles (source of fireballs) 
are unlikely to be released.

\subsubsection{Sublimation of Water Ice}
\label{ice}

Sublimation of water ice may be an improbable 
dust production mechanism for differentiated (V-type) 
or thermally metamorphosed (S-type) asteroids.  
On the contrary, the presence of aqueously altered minerals 
on those of surfaces have been reported \citep[][]{Rivkin15aste,Rivkin18},   
as well as further evidences, 
the weakly active S-type Oort Cloud object driven 
by sublimation of water ice \citep{Meech16} and 
the native water inclusion in Itokawa samples \citep[][]{Jineaav8106}.   
Asteroid (4) Vesta's current surface texture, fracture 
and roughness (cm- to 10\,cm-scale) could be 
caused by (carbonaceous) impactors \citep{Hasegawa03,Sanctis12,Russell13,Russell15}.  
By contrast, the recent Dawn bistatic radar observation indicates 
subsurface volatiles (water ice) involvement processes \citep[][]{Palmer17}. 

Can buried water ice exist and survive even in V-type asteroids?  
The differentiation process would occur for the most part of the body, but partially may not.   
The Vesta's smoother terrain area (heightened hydrogen $>$ 0.015\%),  
on which subsurface ice might contribute to,   
occupy only $\gtrsim$ 0.01\% of the total surface area \citep[][]{Palmer17}.  
The extreme partiality might lead to the localized subsurface ice existence.   

How deep can water ice survive in the 2003~YT$_1$ primary, if it were therein?  
Megaregolith-like materials (large, rubble, brecciated bedrock), 
similar structure found in Vesta \citep{Denevi12,Hoffmann12},  
have low thermal diffusivity $\kappa \sim 10^{-7}-10^{-8}\,{\rm m^2\,s^{-1}}$ \citep{Haack90,Fu14}.   
The diurnal thermal skin depth (at which temperature is reduced to be a factor of $1/e$), 
$d_{\rm s}$, is estimated by $\sim$ $\sqrt{\kappa P_{\rm rot}}$.  
Setting $\kappa$ = 10$^{-7}$--10$^{-8}$\,${\rm m^2\,s^{-1}}$ and 
$P_{\rm rot}$=2.343\,hr find $d_{\rm s}$$\sim$0.9--3\,cm.  
The black body temperature at the thermal skin depth is 
$\sim$120\,K even at perihelion, 
below the sublimation temperature of water ice 150\,K \citep{Yamamoto1985}.
Conceivably, water ice might be preserved 
in the very shallow subsurface within a few cm.

To estimate the size of ejected dust particles coupled to 
the outflowing gas driven by sublimation of water ice, 
the small source approximation (SSA) model is applied \citep[][]{Jewitt14}.  
We assume a small patch of surface water ice on 2003~YT$_1$, 
and also assume that subsurface water ice acts in a similar way to the exposed ice.   
Spacecraft visits to comets find too small 
ice exposure on the nuclei \citep[67P/C-G, ][]{Hu17} 
to explain the measured activities driven by sublimation on which 
a few--10s\,\% of surface ice coverages are presumed to replenish \citep{Tancredi06}.  
Alternatively, shallow subsurface water ice is proposed to the most 
of contribution \citep[67P and Ceres, ][]{Agarwal17MNRAS,Kuppers19}.     
A non-rotating, spherical object is assumed for the physical essence of gas dynamics.  
This prevents complicated gas flows caused by inhomogeneous distribution 
of gas release from the non-spherical object \citep[][]{Fulle15,Agarwal16}. 
Then, ice sublimation from an exposed ($\approx$ subsurface) ice patch located 
at the subsolar point is examined.  
We solved energy balance equation of a completely 
absorbing (sub)surface ice at the subsolar point, 
with 2003~YT$_1$ located at perihelion $q$=0.786\,AU.  
The subsolar temperature at the $d_{\rm s}$ is $\sim$160\,K, 
warm enough for water ice to sublimate.
The flux energy completely absorbed from the sun 
and energy lost from the asteroid surface 
by radiation and latent heat of ice sublimation are calculated. 
The resulting maximum specific mass loss rate is $(dm/dt)_{\rm ice}$ = 8$\times$10$^{-4}$\,kg\,m$^{-2}$\,s$^{-1}$ 
at the subsolar point (at the highest temperature 206\,K of the non-rotating body). 
The terminal velocity in the SSA by gas drag is very small compared to 
the gravitational escape speed from the asteroid, 
but certainly assist to launch dust particles from the surface into interplanetary space. 
The radius of ice sublimating area (patch), $r_{\rm ice}$, 
is related with the critical size of dust particles to be ejected, $a_{\rm c}$, 
as expressed by Equation (A6) of \cite{Jewitt14},    
\begin{equation}
r_{\rm ice} > \frac{8 \pi G \rho^2 D_{\rm p}^2 a_{\rm c}}{9 C_{\rm D} v_{\rm gas}} \left(\frac{dm}{dt}\right)_{\rm ice}^{-1}, 
\label{icepatch}
\end{equation}
where $C_{\rm D} \sim$1 is a dimensionless drag coefficient 
which depends on the shape and nature of the grain  
and $v_{\rm gas}$ is the thermal speed of gas molecules.
We set $a_{\rm c}$=1\,mm--1\,cm 
using 
$v_{\rm gas}$=490\,m\,s$^{-1}$ \citep[Equation (10) of][]{GraykowskiJewitt19}
and 
$(dm/dt)_{\rm ice}$ = 8$\times$10$^{-4}$\,kg\,m$^{-2}$\,s$^{-1}$ 
and again take the same values of $G$, $\rho$ and $D_{\rm p}$. 
We then find $r_{\rm ice} >$ 3--25\,m corresponding to  
the fractional area of (sub)surface ice $\sim$ 0.001--0.05\,\%.  
This value is 10 times smaller than, 
or comparable with those of Vesta \citep[$\gtrsim$ 0.01\%,][]{Palmer17} 
and S-type Oort Cloud object \citep[0.04 to 0.1\%,][]{Meech16}.    
These give a crude but useful estimation, 
by showing that even a tiny (sub)surface 
ice coverage can release the meteoroid-sized particles.  
Yet note that no exposed water ice is observed on 2003~YT$_1$.  
Note also that it is difficult to detect subsurface ice by observations.  
Laboratory data find that even a few mm thickness crust (organic mantle) 
perfectly attenuates the near-infrared absorption 
band depths of the subsurface water ice \citep{Poch16}.  
Spacecraft missions for excavations  
alike 
NASA's Deep Impact \citep{AHearn05,Kasuga06MNRAS,Kasuga07Ad}
and 
JAXA's Hayabusa2 \citep{Watanabe17SSRv} 
would be advantageous for the detection in the km-scale NEAs.  
Until then sublimation of water ice is, at least, remained as 
a potential dust production mechanism for 2003~YT$_1$.

Briefly we have examined a variety of process
capable of launching dust particles from 2003~YT$_1$.  
Rotational instability, impacts 
and radiation pressure 
can product mm to cm-scale dust particles. 
By contrast, resurfacing and photoionization are implausible.   
Insufficient evidence exists in 
thermal fracture and sublimation of ice, going to future work.

\section{Summary} 

We present SonotaCo meteor 
survey of a fireball taken in Japan 
on UT 2017 April 28 at ${\rm 15^{h}\,58^{m}\,19^{s}}$.  
The data is measured for orbit and physical properties.  
Specific detections give the following results.

\begin{enumerate}

\item Radiant point, geocentric velocity and orbital elements of fireball are determined.  
	The similarity to asteroid 2003~YT$_1$ with D-criterions (cf. $D_{\rm SH}$= 0.0079)
	gives an order of smaller values than the significant threshold, 
	indicating a parental association.  

\item Absolute visual magnitude is $M_{\rm v}$=$-$4.10$\pm$0.42\,mag.  
	 Light curves give the meteoroid mass $m$ = 29$\pm$1\,g
	 which corresponds to the size $a_{\rm s}$=2.7$\pm$0.1\,cm 
	 with the density 2700\,kg\,m$^{-3}$.

\item	 Meteor luminous model comprising time 
	 derivative of momentum in drag equation 
	 is suggested to employ a velocity-dependent ablation coefficient, 
	 as determined by $\sigma v^2 <$ 1.  

\item The 2003~YT$_1$ binary could be 
	rotationally disrupted asteroids with mass shedding, consistent with \cite{Pravec07Icar}.    
	The YORP spin-up timescale is $\tau_{\rm Y}$$\sim$2\,Myr,    
	shortly induces rotational instability. 
	The resulting end-state is a breakup/fission 
	if it is the rubble-piled body held by weak cohesive strength $S_{\rm c}$$\sim$240\,N\,m$^{-2}$.  

\item Micrometeorite impactors with $\simeq$\,1\,mm in size 
	sufficiently produce the cm-sized dust particles, 
	given populated near the 2003~YT$_1$ orbit.  

\item Radiation pressure may sweep out 
	the mm-sized particles from 2003~YT$_1$, 
	could be source of faint meteors with apparent magnitude of $\sim$$+$5\,mag.    	 
	The cm-sized particles are too large to be removed.   

\item The other dust production mechanisms are 
	unprovable or pending.  
  
\end{enumerate}


\acknowledgments

We are grateful to SonotaCo for support.  
We appreciate Hideaki Muroishi, Hiroshi Yamakawa, Kazuhiko Yoneguchi, Naoya Saito, 
Hiroyuki Inoue, Chikara Shimoda, Toshio Kamimura and Koji Okano for data contributions. 
We acknowledge Masahisa Yanagisawa, David {\v C}apek, Takaya Okamoto and David Jewitt for discussion, 
and David Asher for review.  
TK gives special thanks to Petr Pravec for presentation source and to Daniel J. Scheeres for scientific editor.  
Finally, we express a deep gratitude to Juraj T{\'o}th and his LOC members of 
Meteoroids 2019 held in Bratislava, Slovakia for providing opportunity to enhance this study.

%

\vspace{5mm}
\facilities{SonotaCo Network, EDMOND}


\software{UFOCaptureHD2, UFOOrbitV2, UFOAnalyzerV2 
\citep{So09, SoSI14,So16,So17}
          }



\appendix




\bibliography{ref}

\begin{thebibliography}{}
\expandafter\ifx\csname natexlab\endcsname\relax\def\natexlab#1{#1}\fi
\providecommand{\url}[1]{\href{#1}{#1}}

\bibitem[{{Abell} {et~al.}(2004){Abell}, {Gaffey}, \&
  {Hardersen}}]{AbellFahhey04}
{Abell}, P.~A., {Gaffey}, M.~J., \& {Hardersen}, P.~S. 2004, in Bulletin of the
  American Astronomical Society, Vol.~36, AAS/Division for Planetary Sciences
  Meeting Abstracts \#36, 1132

\bibitem[{{Abell} {et~al.}(2005){Abell}, {Gaffey}, {Hardersen}, {Vilas},
  {Jarvis}, \& {Landis}}]{AbellGaffey05}
{Abell}, P.~A., {Gaffey}, M.~J., {Hardersen}, P.~S., {et~al.} 2005, in Bulletin
  of the American Astronomical Society, Vol.~37, AAS/Division for Planetary
  Sciences Meeting Abstracts \#37, 627

\bibitem[{{Agarwal} {et~al.}(2013){Agarwal}, {Jewitt}, \&
  {Weaver}}]{Agarwal13ApJ}
{Agarwal}, J., {Jewitt}, D., \& {Weaver}, H. 2013, \apj, 769, 46

\bibitem[{{Agarwal} {et~al.}(2016){Agarwal}, {A'Hearn}, {Vincent},
  {G{\"u}ttler}, {H{\"o}fner}, {Sierks}, {Tubiana}, {Barbieri}, {Lamy},
  {Rodrigo}, {Koschny}, {Rickman}, {Barucci}, {Bertaux}, {Bertini},
  {Boudreault}, {Cremonese}, {Da Deppo}, {Davidsson}, {Debei}, {De Cecco},
  {Deller}, {Fornasier}, {Fulle}, {Gicquel}, {Groussin}, {Guti{\'e}rrez},
  {Hofmann}, {Hviid}, {Ip}, {Jorda}, {Keller}, {Knollenberg}, {Kramm},
  {K{\"u}hrt}, {K{\"u}ppers}, {Lara}, {Lazzarin}, {Lopez Moreno}, {Marzari},
  {Naletto}, {Oklay}, {Shi}, \& {Thomas}}]{Agarwal16}
{Agarwal}, J., {A'Hearn}, M.~F., {Vincent}, J.~B., {et~al.} 2016, \mnras, 462,
  S78

\bibitem[{{Agarwal} {et~al.}(2017){Agarwal}, {Della Corte}, {Feldman},
  {Geiger}, {Merouane}, {Bertini}, {Bodewits}, {Fornasier}, {Gr{\"u}n}, \&
  {Hasselmann}}]{Agarwal17MNRAS}
{Agarwal}, J., {Della Corte}, V., {Feldman}, P.~D., {et~al.} 2017, \mnras, 469,
  s606

\bibitem[{{A'Hearn} {et~al.}(2005){A'Hearn}, {Belton}, {Delamere}, {Kissel},
  {Klaasen}, {McFadden}, {Meech}, {Melosh}, {Schultz}, \&
  {Sunshine}}]{AHearn05}
{A'Hearn}, M.~F., {Belton}, M.~J.~S., {Delamere}, W.~A., {et~al.} 2005,
  Science, 310, 258

\bibitem[{{Asher} {et~al.}(1993){Asher}, {Clube}, \& {Steel}}]{Asher93}
{Asher}, D.~J., {Clube}, S.~V.~M., \& {Steel}, D.~I. 1993, \mnras, 264, 93

\bibitem[{{Binzel} {et~al.}(2010){Binzel}, {Morbidelli}, {Merouane}, {DeMeo},
  {Birlan}, {Vernazza}, {Thomas}, {Rivkin}, {Bus}, \& {Tokunaga}}]{Binzel10}
{Binzel}, R.~P., {Morbidelli}, A., {Merouane}, S., {et~al.} 2010, \nat, 463,
  331

\bibitem[{{Bottke} {et~al.}(1994){Bottke}, {Nolan}, {Greenberg}, \&
  {Kolvoord}}]{Bottke94hdtc}
{Bottke}, W.~F., J., {Nolan}, M.~C., {Greenberg}, R., \& {Kolvoord}, R.~A.
  1994, in Hazards Due to Comets and Asteroids, ed. T.~{Gehrels}, M.~S.
  {Matthews}, \& A.~M. {Schumann} (University of Arizona Press, Tucson),
  337--357

\bibitem[{{Bottke} {et~al.}(2002){Bottke}, {Morbidelli}, {Jedicke}, {Petit},
  {Levison}, {Michel}, \& {Metcalfe}}]{Bottke02}
{Bottke}, W.~F., {Morbidelli}, A., {Jedicke}, R., {et~al.} 2002, \icarus, 156,
  399

\bibitem[{{Bronshten}(1983)}]{Bronshten83}
{Bronshten}, V.~A. 1983, {Physics of meteoric phenomena from the Russian}

\bibitem[{{Brooks}(2006)}]{Brooks06}
{Brooks}, H.~E. 2006, in Bulletin of the American Astronomical Society,
  Vol.~38, American Astronomical Society Meeting Abstracts, 934

\bibitem[{{Brown} {et~al.}(2013){Brown}, {Marchenko}, {Moser}, {Weryk}, \&
  {Cooke}}]{BMM13}
{Brown}, P., {Marchenko}, V., {Moser}, D.~E., {Weryk}, R., \& {Cooke}, W. 2013,
  Meteoritics and Planetary Science, 48, 270

\bibitem[{{Burbine} {et~al.}(2009){Burbine}, {Buchanan}, {Dolkar}, \&
  {Binzel}}]{Burbine09}
{Burbine}, T.~H., {Buchanan}, P.~C., {Dolkar}, T., \& {Binzel}, R.~P. 2009,
  Meteoritics and Planetary Science, 44, 1331

\bibitem[{{Carrillo-S{\'a}nchez} {et~al.}(2015){Carrillo-S{\'a}nchez}, {Plane},
  {Feng}, {Nesvorn{\'y}}, \& {Janches}}]{Carrillo15}
{Carrillo-S{\'a}nchez}, J.~D., {Plane}, J.~M.~C., {Feng}, W., {Nesvorn{\'y}},
  D., \& {Janches}, D. 2015, \grl, 42, 6518

\bibitem[{{Ceplecha} {et~al.}(1998){Ceplecha}, {Borovi{\v c}ka}, {Elford},
  {Revelle}, {Hawkes}, {Porub{\v c}an}, \& {{\v S}imek}}]{CBE98}
{Ceplecha}, Z., {Borovi{\v c}ka}, J., {Elford}, W.~G., {et~al.} 1998, \ssr, 84,
  327

\bibitem[{{Ceplecha} \& {McCrosky}(1976)}]{CeplechaMcCrosky76}
{Ceplecha}, Z., \& {McCrosky}, R.~E. 1976, \jgr, 81, 6257

\bibitem[{{Ceplecha} \& {Revelle}(2005)}]{CR05}
{Ceplecha}, Z., \& {Revelle}, D.~O. 2005, Meteoritics and Planetary Science,
  40, 35

\bibitem[{{Ceplecha} {et~al.}(1993){Ceplecha}, {Spurny}, {Borovicka}, \&
  {Keclikova}}]{Ceplecha93}
{Ceplecha}, Z., {Spurny}, P., {Borovicka}, J., \& {Keclikova}, J. 1993, \aap,
  279, 615

\bibitem[{{Chandler} {et~al.}(2019){Chandler}, {Kueny}, {Gustafsson},
  {Trujillo}, {Robinson}, \& {Trilling}}]{Chandler19}
{Chandler}, C.~O., {Kueny}, J., {Gustafsson}, A., {et~al.} 2019, \apjl, 877,
  L12

\bibitem[{{Chang} {et~al.}(2015){Chang}, {Ip}, {Lin}, {Cheng}, {Ngeow}, {Yang},
  {Waszczak}, {Kulkarni}, {Levitan}, {Sesar}, {Laher}, {Surace}, \&
  {Prince}}]{Chang_Ip_15}
{Chang}, C.-K., {Ip}, W.-H., {Lin}, H.-W., {et~al.} 2015, \apjs, 219, 27

\bibitem[{{Clark} {et~al.}(2019){Clark}, {Wiegert}, \& {Brown}}]{Clark2019}
{Clark}, D.~L., {Wiegert}, P., \& {Brown}, P.~G. 2019, \mnras, 487, L35

\bibitem[{{Cochran} {et~al.}(2004){Cochran}, {Vilas}, {Jarvis}, \&
  {Kelley}}]{Cochran04}
{Cochran}, A.~L., {Vilas}, F., {Jarvis}, K.~S., \& {Kelley}, M.~S. 2004,
  \icarus, 167, 360

\bibitem[{{Cruikshank} {et~al.}(1991){Cruikshank}, {Tholen}, {Hartmann},
  {Bell}, \& {Brown}}]{Cruikshank91}
{Cruikshank}, D.~P., {Tholen}, D.~J., {Hartmann}, W.~K., {Bell}, J.~F., \&
  {Brown}, R.~H. 1991, \icarus, 89, 1

\bibitem[{{Davis} {et~al.}(2002){Davis}, {Durda}, {Marzari}, {Campo Bagatin},
  \& {Gil-Hutton}}]{Davis2002}
{Davis}, D.~R., {Durda}, D.~D., {Marzari}, F., {Campo Bagatin}, A., \&
  {Gil-Hutton}, R. 2002, in Asteroids III, W. F. Bottke Jr., A. Cellino, P.
  Paolicchi, and R. P. Binzel (eds), University of Arizona Press, Tucson,
  p.545-558, 545--558

\bibitem[{{De Sanctis} {et~al.}(2012){De Sanctis}, {Combe}, {Ammannito},
  {Palomba}, {Longobardo}, {McCord}, {Marchi}, {Capaccioni}, {Capria},
  {Mittlefehldt}, {Pieters}, {Sunshine}, {Tosi}, {Zambon}, {Carraro}, {Fonte},
  {Frigeri}, {Magni}, {Raymond}, {Russell}, \& {Turrini}}]{Sanctis12}
{De Sanctis}, M.~C., {Combe}, J.~P., {Ammannito}, E., {et~al.} 2012, \apj, 758,
  L36

\bibitem[{{Delbo} {et~al.}(2011){Delbo}, {Walsh}, {Mueller}, {Harris}, \&
  {Howell}}]{Delbo11}
{Delbo}, M., {Walsh}, K., {Mueller}, M., {Harris}, A.~W., \& {Howell}, E.~S.
  2011, \icarus, 212, 138

\bibitem[{{Denevi} {et~al.}(2012){Denevi}, {Coman}, {Blewett}, {Mittlefehldt},
  {Buczkowski}, {Combe}, {de Sanctis}, {Jaumann}, {Li}, \& {Marchi}}]{Denevi12}
{Denevi}, B.~W., {Coman}, E.~I., {Blewett}, D.~T., {et~al.} 2012, in Lunar and
  Planetary Science Conference, 1943

\bibitem[{{Drummond}(1981)}]{Drummond81}
{Drummond}, J.~D. 1981, \icarus, 45, 545

\bibitem[{{Dumitru} {et~al.}(2017){Dumitru}, {Birlan}, {Popescu}, \&
  {Nedelcu}}]{Dumitru17}
{Dumitru}, B.~A., {Birlan}, M., {Popescu}, M., \& {Nedelcu}, D.~A. 2017, \aap,
  607, A5

\bibitem[{{Fang} \& {Margot}(2012)}]{FangMargot12a}
{Fang}, J., \& {Margot}, J.-L. 2012, \aj, 143, 24

\bibitem[{{Flynn} {et~al.}(2018){Flynn}, {Consolmagno}, {Brown}, \&
  {Macke}}]{Flynn18}
{Flynn}, G.~J., {Consolmagno}, G.~J., {Brown}, P., \& {Macke}, R.~J. 2018,
  Chemie der Erde / Geochemistry, 78, 269

\bibitem[{{Flynn} \& {Durda}(2004)}]{Flynn04}
{Flynn}, G.~J., \& {Durda}, D.~D. 2004, \planss, 52, 1129

\bibitem[{{Fu} {et~al.}(2014){Fu}, {Hager}, {Ermakov}, \& {Zuber}}]{Fu14}
{Fu}, R.~R., {Hager}, B.~H., {Ermakov}, A.~I., \& {Zuber}, M.~T. 2014, \icarus,
  240, 133

\bibitem[{{Fulle} {et~al.}(2015){Fulle}, {Ivanovski}, {Bertini}, {Gutierrez},
  {Lara}, {Sierks}, {Zakharov}, {Della Corte}, {Rotundi}, {Barbieri}, {Lamy},
  {Rodrigo}, {Koschny}, {Rickman}, {Keller}, {Agarwal}, {A'Hearn}, {Barucci},
  {Bertaux}, {Bodewits}, {Cremonese}, {Da Deppo}, {Davidsson}, {Debei}, {De
  Cecco}, {Fornasier}, {Groussin}, {G{\"u}ttler}, {Hviid}, {Ip}, {Jorda},
  {Knollenberg}, {Kramm}, {K{\"u}hrt}, {K{\"u}ppers}, {Lazzarin},
  {Lopez-Moreno}, {Marzari}, {Michalik}, {Naletto}, {Oklay}, {Sabau}, {Thomas},
  {Tubiana}, {Vincent}, \& {Wenzel}}]{Fulle15}
{Fulle}, M., {Ivanovski}, S.~L., {Bertini}, I., {et~al.} 2015, \aap, 583, A14

\bibitem[{{Fulvio} {et~al.}(2016){Fulvio}, {Perna}, {Ieva}, {Brunetto},
  {Kanuchova}, {Blanco}, {Strazzulla}, \& {Dotto}}]{Fulvio16}
{Fulvio}, D., {Perna}, D., {Ieva}, S., {et~al.} 2016, \mnras, 455, 584

\bibitem[{{Gal{\'a}d} {et~al.}(2004){Gal{\'a}d}, {Gajdo{\v{s}}},
  {Korno{\v{s}}}, {Vilagi}, \& {Pravec}}]{Galad04IAU}
{Gal{\'a}d}, A., {Gajdo{\v{s}}}, S., {Korno{\v{s}}}, L., {Vilagi}, J., \&
  {Pravec}, P. 2004, International Astronomical Union Circular, 8336, 4

\bibitem[{{Galiazzo} {et~al.}(2017){Galiazzo}, {Silber}, \&
  {Bancelin}}]{Galiazzo17}
{Galiazzo}, M.~A., {Silber}, E.~A., \& {Bancelin}, D. 2017, Astronomische
  Nachrichten, 338, 375

\bibitem[{{Goldreich} \& {Sari}(2009)}]{Goldreich09}
{Goldreich}, P., \& {Sari}, R. 2009, \apj, 691, 54

\bibitem[{{Graves} {et~al.}(2018){Graves}, {Minton}, {Hirabayashi}, {DeMeo}, \&
  {Carry}}]{Graves18}
{Graves}, K.~J., {Minton}, D.~A., {Hirabayashi}, M., {DeMeo}, F.~E., \&
  {Carry}, B. 2018, \icarus, 304, 162

\bibitem[{{Graves} {et~al.}(2019){Graves}, {Minton}, {Molaro}, \&
  {Hirabayashi}}]{Graves19}
{Graves}, K.~J., {Minton}, D.~A., {Molaro}, J.~L., \& {Hirabayashi}, M. 2019,
  \icarus, 322, 1

\bibitem[{{Graykowski} \& {Jewitt}(2019)}]{GraykowskiJewitt19}
{Graykowski}, A., \& {Jewitt}, D. 2019, \aj, 158, 112

\bibitem[{{Haack} {et~al.}(1990){Haack}, {Rasmussen}, \& {Warren}}]{Haack90}
{Haack}, H., {Rasmussen}, K.~L., \& {Warren}, P.~H. 1990, \jgr, 95, 5111

\bibitem[{{Hasegawa} {et~al.}(2003){Hasegawa}, {Murakawa}, {Ishiguro},
  {Nonaka}, {Takato}, {Davis}, {Ueno}, \& {Hiroi}}]{Hasegawa03}
{Hasegawa}, S., {Murakawa}, K., {Ishiguro}, M., {et~al.} 2003, \grl, 30, 2123

\bibitem[{{Hicks} {et~al.}(2009){Hicks}, {Somers}, {Barajas}, {Foster},
  {McAuley}, \& {Shitanishi}}]{Hicks09}
{Hicks}, M., {Somers}, J., {Barajas}, T., {et~al.} 2009, The Astronomer's
  Telegram, 2289

\bibitem[{{Hirabayashi} {et~al.}(2014){Hirabayashi}, {Scheeres}, {S{\'a}nchez},
  \& {Gabriel}}]{Hirabayashi14ApJ}
{Hirabayashi}, M., {Scheeres}, D.~J., {S{\'a}nchez}, D.~P., \& {Gabriel}, T.
  2014, \apjl, 789, L12

\bibitem[{{Hoffmann} {et~al.}(2012){Hoffmann}, {Nathues}, {Vincent}, \&
  {Sierks}}]{Hoffmann12}
{Hoffmann}, M., {Nathues}, A., {Vincent}, J.~B., \& {Sierks}, H. 2012, in EGU
  General Assembly Conference Abstracts, Vol.~14, 5530

\bibitem[{{Housen} \& {Holsapple}(2011)}]{HousenHolsa11}
{Housen}, K.~R., \& {Holsapple}, K.~A. 2011, \icarus, 211, 856

\bibitem[{{Hu} {et~al.}(2017){Hu}, {Shi}, {Sierks}, {Blum}, {Oberst}, {Fulle},
  {K{\"u}hrt}, {G{\"u}ttler}, {Gundlach}, \& {Keller}}]{Hu17}
{Hu}, X., {Shi}, X., {Sierks}, H., {et~al.} 2017, \mnras, 469, S295

\bibitem[{{Hui} {et~al.}(2019){Hui}, {Kim}, \& {Gao}}]{Hui19}
{Hui}, M.-T., {Kim}, Y., \& {Gao}, X. 2019, \mnras, 488, L143

\bibitem[{{Hui} \& {Li}(2017)}]{Hui17}
{Hui}, M.-T., \& {Li}, J. 2017, \aj, 153, 23

\bibitem[{{Ieva} {et~al.}(2016){Ieva}, {Dotto}, {Lazzaro}, {Perna}, {Fulvio},
  \& {Fulchignoni}}]{Ieva16}
{Ieva}, S., {Dotto}, E., {Lazzaro}, D., {et~al.} 2016, \mnras, 455, 2871

\bibitem[{{Jacobson} \& {Scheeres}(2011)}]{Jacobson11}
{Jacobson}, S.~A., \& {Scheeres}, D.~J. 2011, \apj, 736, L19

\bibitem[{{Janches} {et~al.}(2014){Janches}, {Plane}, {Nesvorn{\'y}}, {Feng},
  {Vokrouhlick{\'y}}, \& {Nicolls}}]{Janches14}
{Janches}, D., {Plane}, J.~M.~C., {Nesvorn{\'y}}, D., {et~al.} 2014, \apj, 796,
  41

\bibitem[{{Jeffers} {et~al.}(2001){Jeffers}, {Manley}, {Bailey}, \&
  {Asher}}]{Jeffers01}
{Jeffers}, S.~V., {Manley}, S.~P., {Bailey}, M.~E., \& {Asher}, D.~J. 2001,
  \mnras, 327, 126

\bibitem[{{Jenniskens}(2017)}]{Jenniskens17}
{Jenniskens}, P. 2017, \planss, 143, 116

\bibitem[{{Jenniskens} \& {Lyytinen}(2005)}]{PJL05}
{Jenniskens}, P., \& {Lyytinen}, E. 2005, \aj, 130, 1286

\bibitem[{{Jewitt}(2012)}]{Jewitt12}
{Jewitt}, D. 2012, \aj, 143, 66

\bibitem[{{Jewitt}(2013)}]{Jewitt13}
---. 2013, \aj, 145, 133

\bibitem[{{Jewitt} {et~al.}(2014{\natexlab{a}}){Jewitt}, {Agarwal}, {Li},
  {Weaver}, {Mutchler}, \& {Larson}}]{Jewitt14R3}
{Jewitt}, D., {Agarwal}, J., {Li}, J., {et~al.} 2014{\natexlab{a}}, \apjl, 784,
  L8

\bibitem[{{Jewitt} {et~al.}(2017){Jewitt}, {Agarwal}, {Li}, {Weaver},
  {Mutchler}, \& {Larson}}]{Jewitt17R3}
---. 2017, \aj, 153, 223

\bibitem[{{Jewitt} {et~al.}(2019{\natexlab{a}}){Jewitt}, {Asmus}, {Yang}, \&
  {Li}}]{Jewitt19Thermal}
{Jewitt}, D., {Asmus}, D., {Yang}, B., \& {Li}, J. 2019{\natexlab{a}}, \aj,
  157, 193

\bibitem[{{Jewitt} {et~al.}(2015){Jewitt}, {Hsieh}, \& {Agarwal}}]{JHA15}
{Jewitt}, D., {Hsieh}, H., \& {Agarwal}, J. 2015, in Asteroids IV, ed.
  P.~{Michel}, F.~E. {DeMeo}, \& W.~F. {Bottke} (University of Arizona Press),
  221--241

\bibitem[{{Jewitt} {et~al.}(2013){Jewitt}, {Ishiguro}, \&
  {Agarwal}}]{Jewitt13ApJ}
{Jewitt}, D., {Ishiguro}, M., \& {Agarwal}, J. 2013, \apjl, 764, L5

\bibitem[{{Jewitt} {et~al.}(2014{\natexlab{b}}){Jewitt}, {Ishiguro}, {Weaver},
  {Agarwal}, {Mutchler}, \& {Larson}}]{Jewitt14}
{Jewitt}, D., {Ishiguro}, M., {Weaver}, H., {et~al.} 2014{\natexlab{b}}, \aj,
  147, 117

\bibitem[{{Jewitt} {et~al.}(2019{\natexlab{b}}){Jewitt}, {Kim}, {Luu},
  {Rajagopal}, {Kotulla}, {Ridgway}, \& {Liu}}]{Jewitt19}
{Jewitt}, D., {Kim}, Y., {Luu}, J., {et~al.} 2019{\natexlab{b}}, \apjl, 876,
  L19

\bibitem[{{Jewitt} \& {Li}(2010)}]{JL10}
{Jewitt}, D., \& {Li}, J. 2010, \aj, 140, 1519

\bibitem[{{Jewitt} {et~al.}(2018){Jewitt}, {Mutchler}, {Agarwal}, \&
  {Li}}]{Jewitt18HST}
{Jewitt}, D., {Mutchler}, M., {Agarwal}, J., \& {Li}, J. 2018, \aj, 156, 238

\bibitem[{{Jewitt} {et~al.}(2010){Jewitt}, {Weaver}, {Agarwal}, {Mutchler}, \&
  {Drahus}}]{Jewitt10Nat}
{Jewitt}, D., {Weaver}, H., {Agarwal}, J., {Mutchler}, M., \& {Drahus}, M.
  2010, \nat, 467, 817

\bibitem[{Jin \& Bose(2019)}]{Jineaav8106}
Jin, Z., \& Bose, M. 2019, Science Advances, 5, doi:10.1126/sciadv.aav8106

\bibitem[{{Jutzi} {et~al.}(2010){Jutzi}, {Michel}, {Benz}, \&
  {Richardson}}]{Jutzi10}
{Jutzi}, M., {Michel}, P., {Benz}, W., \& {Richardson}, D.~C. 2010, \icarus,
  207, 54

\bibitem[{{Kasuga} \& {Jewitt}(2019)}]{KasugaJewitt19}
{Kasuga}, T., \& {Jewitt}, D. 2019, in Meteoroids: Sources of Meteors on Earth
  and Beyond, Ryabova G. O., Asher D. J., and Campbell-Brown M. D. (eds.),
  Cambridge, UK: Cambridge University Press, 336 pp., ISBN 9781108426718, 2019,
  p. 187-209, 187

\bibitem[{{Kasuga} {et~al.}(2007){Kasuga}, {Sato}, \& {Watanabe}}]{Kasuga07Ad}
{Kasuga}, T., {Sato}, M., \& {Watanabe}, J. 2007, Advances in Space Research,
  40, 215

\bibitem[{{Kasuga} {et~al.}(2006){Kasuga}, {Watanabe}, \&
  {Sato}}]{Kasuga06MNRAS}
{Kasuga}, T., {Watanabe}, J.-I., \& {Sato}, M. 2006, \mnras, 373, 1107

\bibitem[{{Kleyna} {et~al.}(2019){Kleyna}, {Hainaut}, {Meech}, {Hsieh},
  {Fitzsimmons}, {Micheli}, {Keane}, {Denneau}, {Tonry}, \&
  {Heinze}}]{Kleyna19}
{Kleyna}, J.~T., {Hainaut}, O.~R., {Meech}, K.~J., {et~al.} 2019, \apjl, 874,
  L20

\bibitem[{{Korno{\v{s}}} {et~al.}(2014{\natexlab{a}}){Korno{\v{s}}}, {Koukal},
  {Piffl}, \& {T{\'o}th}}]{Korno14pim3}
{Korno{\v{s}}}, L., {Koukal}, J., {Piffl}, R., \& {T{\'o}th}, J.
  2014{\natexlab{a}}, in Proceedings of the International Meteor Conference,
  Poznan, Poland, 22-25 August 2013, ed. M.~{Gyssens}, P.~{Roggemans}, \&
  P.~{Zoladek}, 23--25

\bibitem[{{Korno{\v{s}}} {et~al.}(2014{\natexlab{b}}){Korno{\v{s}}},
  {Matlovi{\v{c}}}, {Rudawska}, {T{\'o}th}, {Hajdukov{\'a}}, {Koukal}, \&
  {Piffl}}]{Korno14me}
{Korno{\v{s}}}, L., {Matlovi{\v{c}}}, P., {Rudawska}, R., {et~al.}
  2014{\natexlab{b}}, in Meteoroids 2013, ed. T.~J. {Jopek}, F.~J.~M.
  {Rietmeijer}, J.~{Watanabe}, \& I.~P. {Williams}, 225--233

\bibitem[{{K{\"u}ppers}(2019)}]{Kuppers19}
{K{\"u}ppers}, M. 2019, Journal of Geophysical Research (Planets), 124, 205

\bibitem[{{Larson} {et~al.}(2004){Larson}, {Grauer}, {Beshore}, {Christensen},
  {Pravec}, {Kaasalainen}, {Nolan}, {Howell}, {Hine}, {Galad}, {Gajdos},
  {Kornos}, \& {Vilagi}}]{Larson04}
{Larson}, S.~M., {Grauer}, A.~D., {Beshore}, E., {et~al.} 2004, in Bulletin of
  the American Astronomical Society, Vol.~36, AAS/Division for Planetary
  Sciences Meeting Abstracts \#36, 1139

\bibitem[{Lauriello(1974)}]{Lauriello1974}
Lauriello, P. 1974, International Journal of Rock Mechanics and Mining Sciences
  \& Geomechanics Abstracts, 11, 75

\bibitem[{{Lindblad}(1987)}]{Lindblad87}
{Lindblad}, B.~A. 1987, in The Evolution of the Small Bodies of the Solar
  System, ed. M.~{Fulchignoni} \& L.~{Kresak}, 229

\bibitem[{{Madiedo} {et~al.}(2014){Madiedo}, {Trigo-Rodr{\'\i}guez}, {Ortiz},
  {Castro-Tirado}, \& {Cabrera-Ca{\~n}o}}]{Madiedo14MNRAS}
{Madiedo}, J.~M., {Trigo-Rodr{\'\i}guez}, J.~M., {Ortiz}, J.~L.,
  {Castro-Tirado}, A.~J., \& {Cabrera-Ca{\~n}o}, J. 2014, \mnras, 443, 1643

\bibitem[{{Madiedo} {et~al.}(2013){Madiedo}, {Trigo-Rodr{\'\i}guez},
  {Williams}, {Ortiz}, \& {Cabrera}}]{Madiedo13MNRAS}
{Madiedo}, J.~M., {Trigo-Rodr{\'\i}guez}, J.~M., {Williams}, I.~P., {Ortiz},
  J.~L., \& {Cabrera}, J. 2013, \mnras, 431, 2464

\bibitem[{{Mainzer} {et~al.}(2011){Mainzer}, {Grav}, {Bauer}, {Masiero},
  {McMillan}, {Cutri}, {Walker}, {Wright}, {Eisenhardt}, {Tholen}, {Spahr},
  {Jedicke}, {Denneau}, {DeBaun}, {Elsbury}, {Gautier}, {Gomillion}, {Hand},
  {Mo}, {Watkins}, {Wilkins}, {Bryngelson}, {Del Pino Molina}, {Desai},
  {G{\'o}mez Camus}, {Hidalgo}, {Konstantopoulos}, {Larsen}, {Maleszewski},
  {Malkan}, {Mauduit}, {Mullan}, {Olszewski}, {Pforr}, {Saro}, {Scotti}, \&
  {Wasserman}}]{Mainzer11}
{Mainzer}, A., {Grav}, T., {Bauer}, J., {et~al.} 2011, \apj, 743, 156

\bibitem[{{Mainzer} {et~al.}(2012){Mainzer}, {Grav}, {Masiero}, {Bauer},
  {Cutri}, {McMillan}, {Nugent}, {Tholen}, {Walker}, \& {Wright}}]{Mainzer12}
{Mainzer}, A., {Grav}, T., {Masiero}, J., {et~al.} 2012, \apj, 760, L12

\bibitem[{{Margot} {et~al.}(2015){Margot}, {Pravec}, {Taylor}, {Carry}, \&
  {Jacobson}}]{Margot15aste}
{Margot}, J.~L., {Pravec}, P., {Taylor}, P., {Carry}, B., \& {Jacobson}, S.
  2015, in Asteroids IV, Patrick Michel, Francesca E. DeMeo, and William F.
  Bottke (eds.), University of Arizona Press, Tucson, 895 pp. ISBN:
  978-0-816-53213-1, 2015., p.355-374, 355--374

\bibitem[{{Marsset} {et~al.}(2019){Marsset}, {DeMeo}, {Sonka}, {Birlan},
  {Polishook}, {Burt}, {Binzel}, {Bus}, \& {Thomas}}]{Marsset19}
{Marsset}, M., {DeMeo}, F., {Sonka}, A., {et~al.} 2019, \apjl, 882, L2

\bibitem[{{Meech} {et~al.}(2016){Meech}, {Yang}, {Kleyna}, {Hainaut},
  {Berdyugina}, {Keane}, {Micheli}, {Morbidelli}, \& {Wainscoat}}]{Meech16}
{Meech}, K.~J., {Yang}, B., {Kleyna}, J., {et~al.} 2016, Science Advances, 2,
  e1600038

\bibitem[{{Molaro} {et~al.}(2015){Molaro}, {Byrne}, \& {Langer}}]{Molaro15}
{Molaro}, J.~L., {Byrne}, S., \& {Langer}, S.~A. 2015, Journal of Geophysical
  Research (Planets), 120, 255

\bibitem[{{Morbidelli} {et~al.}(2002){Morbidelli}, {Bottke}, {Froeschl{\'e}},
  \& {Michel}}]{Morbidelli02aste}
{Morbidelli}, A., {Bottke}, W.~F., J., {Froeschl{\'e}}, C., \& {Michel}, P.
  2002, in Asteroids III, W. F. Bottke Jr., A. Cellino, P. Paolicchi, and R. P.
  Binzel (eds), University of Arizona Press, Tucson, p.409-422, 409--422

\bibitem[{{Moreno} {et~al.}(2019){Moreno}, {Jehin}, {Licandro}, {Ferrais},
  {Moulane}, {Pozuelos}, {Manfroid}, {Devog{\`e}le}, {Benkhaldoun}, \&
  {Moskovitz}}]{Moreno19}
{Moreno}, F., {Jehin}, E., {Licandro}, J., {et~al.} 2019, \aap, 624, L14

\bibitem[{{Moskovitz} {et~al.}(2019){Moskovitz}, {Fatka}, {Farnocchia},
  {Devog{\`e}le}, {Polishook}, {Thomas}, {Mommert}, {Avner}, {Binzel}, {Burt},
  {Christensen}, {DeMeo}, {Hinkle}, {Hora}, {Magnusson}, {Matson}, {Person},
  {Skiff}, {Thirouin}, {Trilling}, {Wasserman}, \& {Willman}}]{Moskovitz19}
{Moskovitz}, N.~A., {Fatka}, P., {Farnocchia}, D., {et~al.} 2019, \icarus, 333,
  165

\bibitem[{{Myers} {et~al.}(2001){Myers}, {Sande}, {Miller}, {Warren}, \&
  {Tracewell}}]{Myers01}
{Myers}, J.~R., {Sande}, C.~B., {Miller}, A.~C., {Warren}, W.~H., J., \&
  {Tracewell}, D.~A. 2001, VizieR Online Data Catalog, V/109

\bibitem[{{Nagasawa}(1981)}]{Nagasawa81}
{Nagasawa}, K. 1981, The Earth Monthly, 3, 588

\bibitem[{{Neslusan} {et~al.}(1998){Neslusan}, {Svoren}, \&
  {Porubcan}}]{Neslusan98}
{Neslusan}, L., {Svoren}, J., \& {Porubcan}, V. 1998, \aap, 331, 411

\bibitem[{{Nesvorn{\'y}} {et~al.}(2010){Nesvorn{\'y}}, {Jenniskens}, {Levison},
  {Bottke}, {Vokrouhlick{\'y}}, \& {Gounelle}}]{Nesvorn10}
{Nesvorn{\'y}}, D., {Jenniskens}, P., {Levison}, H.~F., {et~al.} 2010, \apj,
  713, 816

\bibitem[{{Nolan} {et~al.}(2004{\natexlab{a}}){Nolan}, {Howell}, \&
  {Hine}}]{Nolan04IAU}
{Nolan}, M.~C., {Howell}, E.~S., \& {Hine}, A.~A. 2004{\natexlab{a}}, \iaucirc,
  8336

\bibitem[{{Nolan} {et~al.}(2004{\natexlab{b}}){Nolan}, {Howell}, \&
  {Miranda}}]{Nolan04}
{Nolan}, M.~C., {Howell}, E.~S., \& {Miranda}, G. 2004{\natexlab{b}}, in
  Bulletin of the American Astronomical Society, Vol.~36, AAS/Division for
  Planetary Sciences Meeting Abstracts \#36, 1132

\bibitem[{{Olech} {et~al.}(2017){Olech}, {{\.Z}o{\l}{\c a}dek},
  {Wi{\'s}niewski}, {Tymi{\'n}ski}, {Stolarz}, {B{\c e}ben}, {Dorosz},
  {Fajfer}, {Fietkiewicz}, {Gawro{\'n}ski}, {Gozdalski}, {Ka{\l}u{\.z}ny},
  {Krasnowski}, {Krygiel}, {Krzy{\.z}anowski}, {Kwinta}, {{\L}ojek},
  {Maciejewski}, {Miernicki}, {Myszkiewicz}, {Nowak}, {Polak}, {Polakowski},
  {Laskowski}, {Szlagor}, {Tissler}, {Suchodolski}, {W{\c e}grzyk},
  {Wo{\'z}niak}, \& {Zar{\c e}ba}}]{Olech17}
{Olech}, A., {{\.Z}o{\l}{\c a}dek}, P., {Wi{\'s}niewski}, M., {et~al.} 2017,
  \mnras, 469, 2077

\bibitem[{{Palmer} {et~al.}(2017){Palmer}, {Heggy}, \& {Kofman}}]{Palmer17}
{Palmer}, E.~M., {Heggy}, E., \& {Kofman}, W. 2017, Nature Communications, 8,
  409

\bibitem[{{Pariseau}(2006)}]{Pariseau2006}
{Pariseau}, G.~W. 2006, {Design Analysis in Rock Mechanics} (London: Taylor and
  Francis), 576

\bibitem[{{Pieters} {et~al.}(2012){Pieters}, {Ammannito}, {Blewett}, {Denevi},
  {de Sanctis}, {Gaffey}, {Le Corre}, {Li}, {Marchi}, {McCord}, {McFadden},
  {Mittlefehldt}, {Nathues}, {Palmer}, {Reddy}, {Raymond}, \&
  {Russell}}]{Pieters12}
{Pieters}, C.~M., {Ammannito}, E., {Blewett}, D.~T., {et~al.} 2012, \nat, 491,
  79

\bibitem[{{Poch} {et~al.}(2016){Poch}, {Pommerol}, {Jost}, {Carrasco}, {Szopa},
  \& {Thomas}}]{Poch16}
{Poch}, O., {Pommerol}, A., {Jost}, B., {et~al.} 2016, \icarus, 267, 154

\bibitem[{{Popova} {et~al.}(2019){Popova}, {Borovi{\v{c}}ka}, \&
  {Campbell-Brown}}]{Popova19}
{Popova}, O., {Borovi{\v{c}}ka}, J., \& {Campbell-Brown}, M.~D. 2019, in
  Meteoroids: Sources of Meteors on Earth and Beyond, Ryabova G. O., Asher D.
  J., and Campbell-Brown M. D. (eds.), Cambridge, UK: Cambridge University
  Press, 336 pp., ISBN 9781108426718, 2019, p. 9-36, 9

\bibitem[{{Pravec}(2005)}]{Pravec05SASS}
{Pravec}, P. 2005, Society for Astronomical Sciences Annual Symposium, 24, 61

\bibitem[{{Pravec} \& {Harris}(2007)}]{Pravec07Icar}
{Pravec}, P., \& {Harris}, A.~W. 2007, \icarus, 190, 250

\bibitem[{{Pravec} {et~al.}(2008){Pravec}, {Harris}, {Vokrouhlick{\'y}},
  {Warner}, {Ku{\v{s}}nir{\'a}k}, {Hornoch}, {Pray}, {Higgins}, {Oey},
  {Gal{\'a}d}, {Gajdo{\v{s}}}, {Korno{\v{s}}}, {Vil{\'a}gi}, {Hus{\'a}rik},
  {Krugly}, {Shevchenko}, {Chiorny}, {Gaftonyuk}, {Cooney}, {Gross}, {Terrell},
  {Stephens}, {Dyvig}, {Reddy}, {Ries}, {Colas}, {Lecacheux}, {Durkee}, {Masi},
  {Koff}, \& {Goncalves}}]{Pravec08Icar}
{Pravec}, P., {Harris}, A.~W., {Vokrouhlick{\'y}}, D., {et~al.} 2008, \icarus,
  197, 497

\bibitem[{{Pravec} {et~al.}(2016){Pravec}, {Scheirich}, {Ku{\v s}nir{\'a}k},
  {Hornoch}, {Gal{\'a}d}, {Naidu}, {Pray}, {Vil{\'a}gi}, {Gajdo{\v s}},
  {Korno{\v s}}, {Krugly}, {Cooney}, {Gross}, {Terrell}, {Gaftonyuk},
  {Pollock}, {Hus{\'a}rik}, {Chiorny}, {Stephens}, {Durkee}, {Reddy}, {Dyvig},
  {Vra{\v s}til}, {{\v Z}i{\v z}ka}, {Mottola}, {Hellmich}, {Oey}, {Benishek},
  {Kryszczy{\'n}ska}, {Higgins}, {Ries}, {Marchis}, {Baek}, {Macomber},
  {Inasaridze}, {Kvaratskhelia}, {Ayvazian}, {Rumyantsev}, {Masi}, {Colas},
  {Lecacheux}, {Montaigut}, {Leroy}, {Brown}, {Krzeminski}, {Molotov},
  {Reichart}, {Haislip}, \& {LaCluyze}}]{Pravec16}
{Pravec}, P., {Scheirich}, P., {Ku{\v s}nir{\'a}k}, P., {et~al.} 2016, \icarus,
  267, 267

\bibitem[{Richter \& Simmons(1974)}]{Richter1974}
Richter, D., \& Simmons, G. 1974, International Journal of Rock Mechanics and
  Mining Sciences \& Geomechanics Abstracts, 11, 403

\bibitem[{{Rivkin} {et~al.}(2015){Rivkin}, {Campins}, {Emery}, {Howell},
  {Licandro}, {Takir}, \& {Vilas}}]{Rivkin15aste}
{Rivkin}, A.~S., {Campins}, H., {Emery}, J.~P., {et~al.} 2015, in Asteroids IV,
  Patrick Michel, Francesca E. DeMeo, and William F. Bottke (eds.), University
  of Arizona Press, Tucson, 895 pp. ISBN: 978-0-816-53213-1, 2015., p.65-87,
  65--87

\bibitem[{{Rivkin} {et~al.}(2018){Rivkin}, {Howell}, {Emery}, \&
  {Sunshine}}]{Rivkin18}
{Rivkin}, A.~S., {Howell}, E.~S., {Emery}, J.~P., \& {Sunshine}, J. 2018,
  \icarus, 304, 74

\bibitem[{{Rozitis} \& {Green}(2013)}]{RozitisGreen13}
{Rozitis}, B., \& {Green}, S.~F. 2013, \mnras, 430, 1376

\bibitem[{{Rudawska} \& {Jenniskens}(2014)}]{RudawskaJenniskens2014}
{Rudawska}, R., \& {Jenniskens}, P. 2014, in Meteoroids 2013, ed. T.~J.
  {Jopek}, F.~J.~M. {Rietmeijer}, J.~{Watanabe}, \& I.~P. {Williams}, 217--224

\bibitem[{{Russell} {et~al.}(2015){Russell}, {McSween}, {Jaumann}, \&
  {Raymond}}]{Russell15}
{Russell}, C.~T., {McSween}, H.~Y., {Jaumann}, R., \& {Raymond}, C.~A. 2015, in
  Asteroids IV, Patrick Michel, Francesca E. DeMeo, and William F. Bottke
  (eds.), University of Arizona Press, Tucson, 895 pp. ISBN: 978-0-816-53213-1,
  2015., p.419-432, 419--432

\bibitem[{{Russell} {et~al.}(2013){Russell}, {Raymond}, {Jaumann}, {McSween},
  {De Sanctis}, {Nathues}, {Prettyman}, {Ammannito}, {Reddy}, {Preusker},
  {O'Brien}, {Marchi}, {Denevi}, {Buczkowski}, {Pieters}, {McCord}, {Li},
  {Mittlefehldt}, {Combe}, {Williams}, {Hiesinger}, {Yingst}, {Polanskey}, \&
  {Joy}}]{Russell13}
{Russell}, C.~T., {Raymond}, C.~A., {Jaumann}, R., {et~al.} 2013, Meteoritics
  and Planetary Science, 48, 2076

\bibitem[{{Sanchez} {et~al.}(2013){Sanchez}, {Michelsen}, {Reddy}, \&
  {Nathues}}]{Sanchez13}
{Sanchez}, J.~A., {Michelsen}, R., {Reddy}, V., \& {Nathues}, A. 2013, \icarus,
  225, 131

\bibitem[{{Sato} \& {Watanabe}(2014)}]{Sato_Watanabe14}
{Sato}, M., \& {Watanabe}, J. 2014, Meteoroids 2013, 329

\bibitem[{{Scheeres} {et~al.}(2010){Scheeres}, {Hartzell}, {S{\'a}nchez}, \&
  {Swift}}]{Scheeres10}
{Scheeres}, D.~J., {Hartzell}, C.~M., {S{\'a}nchez}, P., \& {Swift}, M. 2010,
  \icarus, 210, 968

\bibitem[{{Scheeres} \& {S{\'a}nchez}(2018)}]{Scheeres18}
{Scheeres}, D.~J., \& {S{\'a}nchez}, P. 2018, Progress in Earth and Planetary
  Science, 5, 25

\bibitem[{{SonotaCo}(2009)}]{So09}
{SonotaCo}. 2009, WGN, Journal of the International Meteor Organization, 37, 55

\bibitem[{{SonotaCo}(2016)}]{So16}
---. 2016, WGN, Journal of the International Meteor Organization, 44, 42

\bibitem[{{SonotaCo}(2017)}]{So17}
---. 2017, WGN, Journal of the International Meteor Organization, 45, 95

\bibitem[{{SonotaCo} {et~al.}(2014){SonotaCo}, {Shimoda}, {Inoue}, {Masuzawa},
  \& {Sato}}]{SoSI14}
{SonotaCo}, {Shimoda}, C., {Inoue}, H., {Masuzawa}, T., \& {Sato}, M. 2014,
  WGN, Journal of the International Meteor Organization, 42, 222

\bibitem[{{Southworth} \& {Hawkins}(1963)}]{SouthHawkins63}
{Southworth}, R.~B., \& {Hawkins}, G.~S. 1963, Smithsonian Contributions to
  Astrophysics, 7, 261

\bibitem[{{Spurn{\'y}} {et~al.}(2017){Spurn{\'y}}, {Borovi{\v c}ka}, {Mucke},
  \& {Svore{\v n}}}]{SBM17}
{Spurn{\'y}}, P., {Borovi{\v c}ka}, J., {Mucke}, H., \& {Svore{\v n}}, J. 2017,
  \aap, 605, A68

\bibitem[{{Subasinghe} \& {Campbell-Brown}(2018)}]{Subasinghe18}
{Subasinghe}, D., \& {Campbell-Brown}, M. 2018, \aj, 155, 88

\bibitem[{{Svetsov} {et~al.}(2019){Svetsov}, {Shuvalov}, {Collins}, \&
  {Popova}}]{Svetsov19}
{Svetsov}, V., {Shuvalov}, V., {Collins}, G., \& {Popova}, O. 2019, in
  Meteoroids: Sources of Meteors on Earth and Beyond, Ryabova G. O., Asher D.
  J., and Campbell-Brown M. D. (eds.), Cambridge, UK: Cambridge University
  Press, 336 pp., ISBN 9781108426718, 2019, p. 275-298, 275

\bibitem[{{Tancredi} {et~al.}(2006){Tancredi}, {Fern{\'a}ndez}, {Rickman}, \&
  {Licandro}}]{Tancredi06}
{Tancredi}, G., {Fern{\'a}ndez}, J.~A., {Rickman}, H., \& {Licandro}, J. 2006,
  \icarus, 182, 527

\bibitem[{{Tichy} {et~al.}(2003){Tichy}, {Kocer}, {Yeung}, {Kusnirak},
  {Larson}, {Beshore}, {Christensen}, {Hill}, {Kolar}, {McLean}, {McGaha},
  {Holvorcem}, {Schwartz}, \& {Spahr}}]{Tichy03}
{Tichy}, M., {Kocer}, M., {Yeung}, W.~K.~Y., {et~al.} 2003, Minor Planet
  Electronic Circulars, 2003-Y30

\bibitem[{{Tsuchiya} {et~al.}(2017){Tsuchiya}, {Sato}, {Watanabe}, {Moorhead},
  {Moser}, {Brown}, \& {Cooke}}]{Tsuchiya17}
{Tsuchiya}, C., {Sato}, M., {Watanabe}, J.-i., {et~al.} 2017, \planss, 143, 142

\bibitem[{{Vaubaillon} {et~al.}(2019){Vaubaillon}, {Neslu{\v{s}}an}, {Sekhar},
  {Rudawska}, \& {Ryabova}}]{Vaubaillon19}
{Vaubaillon}, J., {Neslu{\v{s}}an}, L., {Sekhar}, A., {Rudawska}, R., \&
  {Ryabova}, G.~O. 2019, {From Parent Body to Meteor Shower: The Dynamics of
  Meteoroid Streams}, 161

\bibitem[{{Vokrouhlick{\'y}} {et~al.}(2015){Vokrouhlick{\'y}}, {Bottke},
  {Chesley}, {Scheeres}, \& {Statler}}]{Vokrouhlick15}
{Vokrouhlick{\'y}}, D., {Bottke}, W.~F., {Chesley}, S.~R., {Scheeres}, D.~J.,
  \& {Statler}, T.~S. 2015, in Asteroids IV, ed. F.~E.~D. Patrick~Michel \&
  W.~F. Bottke (University of Arizona Press, Tucson), 509--531

\bibitem[{{Walsh} \& {Jacobson}(2015)}]{Walsh15aste}
{Walsh}, K.~J., \& {Jacobson}, S.~A. 2015, in Asteroids IV, Patrick Michel,
  Francesca E. DeMeo, and William F. Bottke (eds.), University of Arizona
  Press, Tucson, 895 pp. ISBN: 978-0-816-53213-1, 2015., p.375-393, 375--393

\bibitem[{{Walsh} {et~al.}(2008){Walsh}, {Richardson}, \& {Michel}}]{Walsh08}
{Walsh}, K.~J., {Richardson}, D.~C., \& {Michel}, P. 2008, \nat, 454, 188

\bibitem[{{Warner} {et~al.}(2018){Warner}, {Pravec}, \& {Harris}}]{Warner18PDS}
{Warner}, B., {Pravec}, P., \& {Harris}, A.~P. 2018, NASA Planetary Data System

\bibitem[{{Warner} {et~al.}(2009){Warner}, {Harris}, \& {Pravec}}]{WHP09}
{Warner}, B.~D., {Harris}, A.~W., \& {Pravec}, P. 2009, \icarus, 202, 134

\bibitem[{{Watanabe} {et~al.}(2017){Watanabe}, {Tsuda}, {Yoshikawa}, {Tanaka},
  {Saiki}, \& {Nakazawa}}]{Watanabe17SSRv}
{Watanabe}, S.-i., {Tsuda}, Y., {Yoshikawa}, M., {et~al.} 2017, \ssr, 208, 3

\bibitem[{{Weryk} \& {Brown}(2013)}]{Weryk13pss}
{Weryk}, R.~J., \& {Brown}, P.~G. 2013, \planss, 81, 32

\bibitem[{{Williams} {et~al.}(2019){Williams}, {Jopek}, {Rudawska}, {T{\'o}th},
  \& {Korno{\v{s}}}}]{Williams19}
{Williams}, I.~P., {Jopek}, T.~J., {Rudawska}, R., {T{\'o}th}, J., \&
  {Korno{\v{s}}}, L. 2019, in Meteoroids: Sources of Meteors on Earth and
  Beyond, Ryabova G. O., Asher D. J., and Campbell-Brown M. D. (eds.),
  Cambridge, UK: Cambridge University Press, 336 pp., ISBN 9781108426718, 2019,
  p. 210-234, 210

\bibitem[{{Yamamoto}(1985)}]{Yamamoto1985}
{Yamamoto}, T. 1985, \aap, 142, 31

\bibitem[{{Ye} {et~al.}(2019){Ye}, {Kelley}, {Bodewits}, {Bolin}, {Jones},
  {Lin}, {Bellm}, {Dekany}, {Duev}, \& {Groom}}]{Ye19}
{Ye}, Q., {Kelley}, M. S.~P., {Bodewits}, D., {et~al.} 2019, \apjl, 874, L16

\bibitem[{{Ye} {et~al.}(2016){Ye}, {Brown}, \& {Pokorn{\'y}}}]{Ye2016}
{Ye}, Q.-Z., {Brown}, P.~G., \& {Pokorn{\'y}}, P. 2016, \mnras, 462, 3511

\end{thebibliography}
\bibliographystyle{aasjournal}



\appendix

\section{Meteor Search in SonotaCo and EDMOND Databases}
\label{SonoEd}

We used the SonotaCo and EDMOND databases 
to find other probable meteors which could be orbitally 
associated with asteroid 2003~YT$_1$ and the Kyoto fireball.   
Note that slow-speed meteors infer large uncertainties 
in the radiant points \citep{Sato_Watanabe14,Tsuchiya17}, 
while  
the databases include little or nothing about estimations for errors on orbital information.  
Hence 
based on the asteroidal solar longitudes and radiant points in Table~\ref{radv}, 
we set the wide search ranges 
$\lambda_{\rm s}-$30$^\circ$ $\leqslant$ $\lambda_{\rm s}'$ $\leqslant$ $\lambda_{\rm s}$+30$^\circ$, 
$\alpha -$30$^\circ$ $\leqslant$ $\alpha'$ $\leqslant$ $\alpha$+30$^\circ$, 
$\delta-$20$^\circ$ $\leqslant$ $\delta'$ $\leqslant$ $\delta+$20$^\circ$
and 
$v_{\rm g} <$ 35\,km\,s$^{-1}$ 
at descending or ascending node respectively.   
In which 
$\lambda_{\rm s}'$, $\alpha'$ and $\delta'$ 
are those of meteors in the databases.
Among them we take 
any of D-criterion for either the asteroid or the fireball 
presents the value of $<$ 0.2.     
Selected meteors are shown in Tables~\ref{survey} and \ref{Dsurvey}.  
The scattered results may not be sufficient to be part of the association.

\section{Meteor Luminous Model}
\label{appendixA}

We present a procedure to develop a new meteor luminous model 
based on the classical model \citep[][]{Bronshten83,CBE98} \citep[Reviewed in][]{Popova19}.  
The meteoroid kinetic energy is transformed 
into radiation during the meteor flight.
The classical luminous model equating mass loss (ablation), 
luminosity and deceleration is given by \cite[Chapter~3.4 in][]{CBE98},  
\begin{eqnarray}
I & = & - \tau \frac{d}{dt}  \left(\frac{m v^2}{2} \right) \nonumber \\
\label{basic}
  & = & - \tau \left( \frac{v^2}{2} \frac{dm}{dt} + mv \frac{dv}{dt} \right) \\
\label{I_C98}
  & = & - \tau \left( 1+ \frac{2}{\sigma v^2} \right) \frac{v^2}{2} \frac{dm}{dt}, 
\end{eqnarray}
where $I$ is the meteor luminosity, 
$\tau$ is the luminous efficiency, 
$m$ is the meteoroid mass (g), 
$\sigma$ is the ablation coefficient (s$^{2}$\,km$^{-2}$) (= kg MJ$^{-1}$), 
$v$ is the meteor velocity (cm\,s$^{-1}$) and $t$ (s) is time.
The meteor luminosity is defined as $I = 10^{-0.4 M_{\rm v}}$ 
in magnitude-based units in the visual region, 
where $M_{\rm v}$ is the absolute magnitude in 100\,km distance.  
The ablation coefficient is generally defined as $\sigma = \Lambda / 2Q \Gamma$, 
where $\Lambda$ is the heat transfer coefficient, 
$Q$ is the energy necessary to ablate an unit mass of meteoroid (MJ kg$^{-1}$) 
and $ \Gamma$ is the drag coefficient.  
The motion and ablation of single non-fragmenting body through the atmosphere 
has been traditionally represented by the drag and mass-loss equations as \citep[Chapter~3.2 in][]{CBE98}, 
\begin{equation}
m \frac{dv}{dt}  = - \Gamma S \rho_{\rm a} v^2,
\label{drag_former}
\end{equation}
\begin{equation}
\frac{dm}{dt} = -\frac{\Lambda S}{2Q} \rho_{\rm a} v^3, 
\label{mass-loss}
\end{equation}
respectively.  
Here $S$ is the cross-section of meteoroid and $\rho_{\rm a}$ is the atmospheric density.  
The Equations~(\ref{drag_former}) and (\ref{mass-loss}) are related as 
\begin{equation}
\frac{dv}{dt} =  \frac{1}{\sigma m v} \frac{dm}{dt}.  
\label{dvdt}
\end{equation}
Substitution Equation (\ref{dvdt}) into (\ref{basic}), we obtain the classical luminous model in Equation~(\ref{I_C98}).

A new luminous model is developed by refining the drag equation (\ref{drag_former}). 
Since the ablation process can lose the mass of meteoroid itself, 
the drag force should be expressed in 
differential form of the momentum \citep[][in Japanese]{Nagasawa81}.  
The drag equation~(\ref{drag_former}) thus can be rewritten as    
\begin{equation}
\frac{d}{dt} (mv) = m \frac{dv}{dt} + \frac{dm}{dt}v = - \Gamma S \rho_{\rm a} v^2.   
\label{momentum}
\end{equation}
Substitution Equation (\ref{mass-loss}) into (\ref{momentum}), the refined drag equation is obtained as  
\begin{equation}
m \frac{dv}{dt}  =  - \Gamma S \rho_{\rm a} v^2 (1- \sigma v^2 ),
\label{newdrag}
\end{equation}
where $\sigma v^2 <$ 1 is required. 
The new relation between $dv/dt$ and $dm/dt$ using Equations~(\ref{mass-loss}) and (\ref{newdrag}) finds 
\begin{equation}
\frac{dv}{dt} = \frac{1-\sigma v^2}{\sigma mv} \frac{dm}{dt}.  
\label{newdvdt}
\end{equation}
By substituting Equation~(\ref{newdvdt}) into (\ref{basic}), we obtain the new luminous model 
 \begin{equation}
I   =  - \tau \left(\frac{2}{\sigma v^2} - 1  \right)\frac{v^2}{2} \frac{dm}{dt}.    
\label{newmodel}
\end{equation}

The ablation coefficient, $\sigma$, 
characterizes the ability of meteoroid to ablate.  
A larger value produces higher mass-loss, 
resulting in brighter luminosity.  
The estimated values in the published literature are, 
however, highly scattered and inconclusive. 
The distribution of coefficient (single-body theory) 
showing 0.01 $<$ $\sigma$ $<$ 0.6\,s$^{2}$\,km$^{-2}$ 
is used to classify the meteoroid type, 
such as ordinary (0.014\,s$^{2}$\,km$^{-2}$),  
carbonaceous chondrites (0.042\,s$^{2}$\,km$^{-2}$) 
or soft cometary materials (0.21\,s$^{2}$\,km$^{-2}$) and so on \citep{Ceplecha93}.
Later, on the contrary, fragmentation process 
is suggested to be dominant for mass loss, 
finding the low $\sigma$ =  0.004 to 0.008\,s$^2$\,km$^{-2}$ 
in any type of meteorite \citep{CR05}.   
But the process also depends on the assumed models.  
The works are reviewed in more detail by \cite{Popova19}.

Here, we propose an appropriate ablation coefficient for the new luminous model therein.   
It concisely depends on meteor velocity, as determined by $\sigma v^2 <$ 1 (see Equation~(\ref{newdrag})). 
Setting $v$=23.7\,km\,s$^{-1}$ (fireball) finds $\sigma <$ 0.00178\,s$^{2}$\,km$^{-2}$.  
We thus use $\sigma$= 0.0017\,s$^{2}$\,km$^{-2}$ for this study.


\clearpage
\begin{deluxetable}{lccc}
\tablecaption{The Kyoto Fireball Trajectory \label{tabletrajectory}}
\tablewidth{0pt}
\tablehead{
       &\colhead{Longitude} &\colhead{Latitude} & \colhead{Height}\\
       &        (deg E)             &     (deg N)             &       (km)  }
\startdata
Beginning      & 136.0156$\pm$0.0004 & 35.4275$\pm$0.0005 & 88.80$\pm$0.07\\
End               & 135.4746$\pm$0.0002 & 34.9859$\pm$0.0003 & 47.80$\pm$0.04\\
\enddata
\tablecomments{The atmospheric trajectory for the fireball (UT 2017 April 28) 
are determined by five camera measurements. 
The observing IDs are TK8\_S7, Osaka03\_3N, 
Osaka03\_06\footnote{The location: Osaka03 obtained the fireball data with two cameras which are expressed as Osaka03\_3N and Osaka03\_06 respectively.}, 
IS2\_S and IS5\_SW (see Figure~\ref{map}).  
The orbital properties (e.g the elements), speed and positioning accuracy (cf. section~\ref{sonotaco}) 
give the estimation of uncertainties.}
\end{deluxetable}

\clearpage

\begin{deluxetable*}{lcccccc}
\tablecaption{Radiant Point and Geocentric Velocity \label{radv}}
\tablewidth{0pt}
\tablehead{
\colhead{Object}  &\colhead{$\alpha$\tablenotemark{a}} & \colhead{$\delta$\tablenotemark{b}} & \colhead{$v_{\rm g}$\tablenotemark{c}} &  \colhead{$\Delta r$\tablenotemark{d}}    & \colhead{$\lambda_{\rm s}$\tablenotemark{e}} &   \colhead{UT Date} \\
             &   (deg)                         &           (deg)           &          (km\,s$^{-1}$)             &      (AU)                       &  (deg)                             &           }
\startdata
Fireball     & 321.2$\pm$0.5                   & +51.2$\pm$0.3             &         23.7$\pm$0.5               &   -                             & 38.3150                         & Apr 28.7, 2017  \\
2003~YT$_1$  &    320.54                       &     +50.97                &          23.7                      &   0.0026                        & 38.3333                              & Apr 28.7, 2017  \\
             &    87.30                        &     $-$36.10              &          23.7                      &   0.0279                        & 218.3331                            & Oct 31.8, 2017  \\
\enddata
\tablecomments{
Radiant point and geocentric velocity of 2003~YT$_1$ are 
calculated by the parallel shift of moving vector at each orbital node \citep[method ($P$) in][]{Neslusan98}.  
The descending node is at $\lambda_{\rm s}$=38.3333$^\circ$ (Apr 28.7, 2017) and the ascending node is at $\lambda_{\rm s}$=218.3331$^\circ$ (Oct 31.8, 2017). 
}
\tablenotetext{a}{Right ascension (J2000.0).}
\tablenotetext{b}{Declination (J2000.0).} 
\tablenotetext{c}{Geocentric velocity.}
\tablenotetext{d}{Distance from descending/ascending node to the Earth's orbit.}  
\tablenotetext{e}{Solar longitude (J2000.0).}
\end{deluxetable*}

\clearpage

\begin{deluxetable*}{lccccccccc}
\tablecaption{Orbital Elements and Period \label{orbit}}
\tablewidth{0pt}
\tablehead{
\colhead{Object}  &\colhead{$a$\tablenotemark{a}} & \colhead{$e$\tablenotemark{b}} & \colhead{$i$\tablenotemark{c}} & \colhead{$q$\tablenotemark{d}} & \colhead{$\omega$\tablenotemark{e}} & \colhead{$\Omega$\tablenotemark{f}} & \colhead{$P_{\rm orb}$\tablenotemark{g}} \\
                  &   (AU)              &                      &      (deg)           &    (AU)               &          (deg)            &         (deg)             &    (yr)      }
\startdata
Fireball          & 1.111$\pm$0.016     & 0.297$\pm$0.004      &  43.9$\pm$0.9        &     0.781$\pm$0.007   &     91.2$\pm$2.7          & 38.315$\pm$0.001          & 1.17 \\
2003~YT$_1$&      1.110          & 0.292                &      44.1            &     0.786             &          91.0             & 38.335                    & 1.17 \\
\enddata
\tablecomments{The uncertainties are propagated from those of radiant point and geocentric velocity (Table~\ref{radv}) 
through the Monte-Carlo technique.  Orbital data of 2003~YT$_1$ are obtained from NASA JPL Small-Body Database Browser (2018).}
\tablenotetext{a}{Semimajor axis.}
\tablenotetext{b}{Eccentricity.} 
\tablenotetext{c}{Inclination.} 
\tablenotetext{d}{Perihelion distance.} 
\tablenotetext{e}{Argument of perihelion.} 
\tablenotetext{f}{Longitude of ascending node.}
\tablenotetext{g}{Orbital period.}
\end{deluxetable*}

\clearpage

\clearpage
\startlongtable
. Make sure there is at least one \tablenotemark
\tablenotetext{a}{Observed Date and Time.}
\tablenotetext{b}{Solar longitude (J2000.0).}
\tablenotetext{c}{Right ascension (J2000.0).}
\tablenotetext{d}{Declination (J2000.0).} 
\tablenotetext{e}{Geocentric velocity.}
\tablenotetext{f}{Semimajor axis.}
\tablenotetext{g}{Eccentricity.} 
\tablenotetext{h}{Inclination.} 
\tablenotetext{i}{Perihelion distance.} 
\tablenotetext{j}{Argument of perihelion.} 
\tablenotetext{k}{Longitude of ascending node.}
\end{deluxetable*}


\clearpage

\begin{deluxetable*}{cccccccc}
\tablecaption{D-criterions for Searched Meteors \label{Dsurvey}}
\tablewidth{0pt}
\tablehead{
\colhead{Data No.} & \multicolumn{3}{c}{2003~YT$_1$} & \hspace{0.5cm} & \multicolumn{3}{c}{Fireball} \\ 
& \colhead{$D_{\rm SH}$\tablenotemark{a}} & \colhead{$D'$\tablenotemark{b}}& \colhead{$D_{\rm ACS}$\tablenotemark{c}} & & \colhead{$D_{\rm SH}$\tablenotemark{a}} & \colhead{$D'$\tablenotemark{b}}& \colhead{$D_{\rm ACS}$\tablenotemark{c}} }
\startdata
\multicolumn{1}{l}{Descending Node}\\
\cline{1-1}
\multicolumn{1}{l}{SonotaCo}\\
1 &0.436	&0.177&	0.394& & 0.439&	0.178&	0.397 \\
2 &0.469	&0.260&	0.180& & 0.468&	0.257&	0.177 \\
3 &0.329	&0.243&	0.202& & 0.325&	0.239&	0.197 \\
4 &0.491	&0.212&	0.163& & 0.494&	0.210&	0.160 \\
5 &0.397	&0.173&	0.133& & 0.401&	0.171&	0.132 \\
\multicolumn{1}{l}{EDMOND}\\    
6 &0.315	&0.348&	0.190& & 0.310&	0.342&	0.185 \\
7 &0.334	&0.159&	0.237& & 0.337&	0.162&	0.241 \\
8 &0.350	&0.346&	0.200& & 0.346&	0.340&	0.198 \\
9 &0.342	&0.191&	0.243& & 0.340&	0.192&	0.239 \\
10&0.370	&0.142&	0.283& & 0.375&	0.145&	0.287 \\
11&0.465	&0.177&	0.205& & 0.470&	0.181&	0.208 \\
12&0.382	&0.201&	0.342& & 0.378&	0.195&	0.337 \\
13&0.235	&0.162&	0.176& & 0.236&	0.161&	0.179 \\
14&0.187	&0.277&	0.144& & 0.181&	0.271&	0.140 \\
15&0.336	&0.185&	0.088& & 0.341&	0.193&	0.092 \\
16&0.477	&0.174&	0.325& & 0.482&	0.177&	0.328 \\
17&0.407	&0.175&	0.401& & 0.410&	0.171&	0.403 \\
18&0.332	&0.195&	0.107& & 0.334&	0.199&	0.110 \\
19&0.158	&0.195&	0.075& & 0.154&	0.192&	0.070 \\
20&0.274	&0.169&	0.269& & 0.275&	0.163&	0.270 \\
\hline
\multicolumn{1}{l}{Ascending Node}\\
\cline{1-1}
\multicolumn{1}{l}{SonotaCo}\\
21&0.185	&0.130&	0.092& &	0.179&	0.121&	0.086 \\
22&0.134	&0.091&	0.094& &	0.127&	0.083&	0.088 \\
\enddata
\tablecomments{Significance is set to $D$ $<$ 0.20 \citep[][]{Williams19}.}
\tablenotetext{a}{\cite{SouthHawkins63}}
\tablenotetext{b}{\cite{Drummond81}}
\tablenotetext{c}{\cite{Asher93}}
\end{deluxetable*}

\clearpage


%
\clearpage

\clearpage

\clearpage


\end{document}